\documentclass[aip,reprint]{revtex4-1}

\usepackage{amssymb,amsmath,amsfonts,amsthm,amsopn}
\usepackage{graphicx}
\usepackage{dcolumn}
\usepackage{bm}
\usepackage{algorithm}
\usepackage{algpseudocode}
\usepackage{xcolor}
\usepackage{url}
\usepackage{epstopdf}
\usepackage{bbm}
\usepackage{cleveref}

\newcommand{\1}{\mathbbm{1}}

\begin{document}


\title{An ``opinion reproduction number'' for infodemics in a bounded-confidence content-spreading process on networks} 

\author{Heather Z. Brooks}
\email[]{hzinnbrooks@g.hmc.edu}
\affiliation{Department of Mathematics, Harvey Mudd College}

\author{Mason A. Porter}
\email[]{mason@math.ucla.edu}
\affiliation{Department of Mathematics, University of California, Los Angeles}
\affiliation{Department of Sociology, University of California, Los Angeles}
\affiliation{Santa Fe Institute}

\date{\today}

\begin{abstract}
    We study the spreading dynamics of content on networks.
    To do this, we use a model in which content spreads through a bounded-confidence mechanism.
    In a bounded-confidence model (BCM) of opinion dynamics, the agents of a network have continuous-valued opinions, which they adjust when they interact with agents whose opinions are sufficiently close to theirs. 
    The employed content-spreading model introduces a twist into BCMs by using bounded confidence for the content spread itself. 
    We define an analogue of the basic reproduction number from disease dynamics that we call an \emph{opinion reproduction number}. 
    A critical value of the opinion reproduction number indicates whether or not there is an ``infodemic'' (i.e., a large content-spreading cascade) of content that reflects a particular opinion. By determining this critical value, one can determine whether or not an opinion dies off or propagates widely as a cascade in a population of agents.
    Using configuration-model networks, we quantify the size and shape of content dissemination {by calculating}
    a variety of summary statistics, and we illustrate how network structure and spreading-model parameters affect these statistics.
    We find that content spreads most widely when the agents have {a} large expected mean degree or {a} large receptiveness to content.
    When the {spreading process slightly exceeds the infodemic threshold,}
    there can be longer dissemination trees than when the expected mean degree or receptiveness {are larger}, even though the total number of content shares is smaller.
\end{abstract}


\maketitle 


\begin{quotation}
    Although most content does not spread far on social media, occasionally some piece of content ``goes viral" and rapidly reaches many people~\cite{goel2016structural,juul2022}.
    When content with misinformation or disinformation spreads widely, it has become common to use
    analogies with disease spread {and state} that there is an \emph{infodemic}~\cite{eysenbach2002,briand2021,zielinski2021}, which is a portmanteau of the words ``information" and ``epidemic". 
    In the present paper, we examine infodemics in a model of content spread that uses a bounded-confidence mechanism. 
    We extend the analogy between the spread of content and the spread of infectious diseases by defining an \emph{opinion reproduction number}, which is inspired by the basic reproduction number of disease dynamics \cite{kiss2017mathematics,brauer2019}. 
    By examining whether or not the spreading process
    is below or above a critical opinion reproduction number (i.e., an \emph{infodemic threshold}), one can examine whether content dies out or goes viral. 
    Our investigation complements branching-process approaches that examine whether or not content spread tends to locally magnify or contract with time \cite{gleeson2021}. 
    In our bounded-confidence model of content spread, we quantify the size and shape of content dissemination using a variety of summary statistics, and we illustrate how network structure and spreading-model parameters affect these statistics in configuration-model networks.
\end{quotation}


\section{Introduction}

Given the enormous scale and impact of human interactions on social media, it is critical to study the collective dynamics that arise in these systems~\cite{bak2021}. Content on social media can take many forms, including scientific facts, other forms of information, hyperlinks to Web pages, pictures, memes, and misinformation and disinformation~\cite{guille2013information,kumpel2015,muhammed2022}. Once created, content can then be magnified by human users, bots, and other accounts, creating an online ecosystem that is filled with {misinformation, disinformation, and ``echo chambers"}~\cite{friggeri2014,starbird2019,yang2023,buitrago2024,echo2024}, even with intentions to share only accurate information~\cite{pennycook2021}. Content sharing by people is ubiquitous and important for social connection, and such sharing can help create and reinforce shared understanding~\cite{baek2023}. Most content does not spread to many people, but some online content {does spread} very far (i.e., it can ``go viral'') in large content-spreading cascades \cite{goel2016structural,juul2022}. A variety of other social phenomena, including emotions and behaviors, can {also spread} as ``social contagions'' on networks \cite{aral2009,christakis2013,borge2013,yu2020,prollochs2021}.

When a piece of misinformation or disinformation spreads very widely, it has become reasonably common to state that there is an \emph{infodemic} \cite{eysenbach2002,briand2021,zielinski2021}. 
The World Health Organization (WHO) gives a much more specific definition of the term ``infodemic" ({which is} a portmanteau of the words ``information" and ``epidemic") in the context of infectious diseases \cite{who-infodemic,infodemic-agenda}: ``An infodemic is too much information including false or misleading information in digital and physical environments during a disease outbreak."
Using both the rigid WHO definition and the looser notion of particularly low-quality content going viral, {it is recognized widely that the COVID-19 pandemic has had an accompanying infodemic} \cite{zarocostas2020,cinelli2020,gallotti2020}. 
Analogously to the spread of infectious diseases, ``superspreader" social-media accounts have played an important role in the COVID-19 infodemic \cite{yang2021}.

Developing a thorough understanding of misinformation, disinformation, and their impact requires a broad view of the problem of ``fake news'' that also {entails} proper understanding of misinformation and its effects~\cite{lazer2018,watts2021}. 
This view needs to encompass disinformation (which is intentionally incorrect), misinformation (which is incorrect, but perhaps unintentionally), biased and misleading information (which may not be factually incorrect), and the production and amplification of such content, including by mainstream news organizations~\cite{buitrago2024}. 
There has been much empirical research on misinformation in areas such as computational social science and allied disciplines~\cite{muhammed2022}.
{Importantly, modeling} efforts can {also play a significant} role in mitigating the harmful effects of misinformation and disinformation \cite{rabb2022}. 
In particular, Juul and Ugander \cite{juul2022} suggested that focusing on reducing the ``infectiousness" of information {and} theoretical analyses of spreading processes may be very helpful for limiting the spread of misinformation and disinformation.

One strategy to gain insight into the mechanisms that underlie observations in social-media systems is by studying mathematical models of opinion dynamics on networks~\cite{noorazar2020classical,noorazar2020recent,peralta2022}. 
Opinion models take a variety of forms. 
The nodes of a network represent agents, and the edges between agents indicate social and/or communication ties between agents.
The opinions of the agents can take either discrete values (e.g., $+1$ or $-1$) or continuous values (e.g., in the interval $[-1,1]$, with $-1$ {representing} the most liberal opinion and $+1$ {representing} the most conservative opinion).
When two (or more) adjacent agents interact, one or more of them updates their opinion according to some rule.
One popular type of {opinion model is a \emph{bounded-confidence model} (BCM)}~\cite{lorenz2007continuous,meng2018opinion,bernardo2024}, in which agents have continuous-valued opinions and interacting agents compromise their opinions by some amount if and only if their opinions are sufficiently close to each other. 
There have been numerous studies of BCMs, which have been generalized in many ways. Recent studies have incorporated phenomena such as media outlets with fixed opinions~\cite{brooks2020model}, polyadic interactions (i.e., interactions between three or more agents) \cite{hickok2022bounded, wang2022opinion}, noise \cite{goddard2022noisy}, asymmetric confidence {bounds} \cite{bernardo2022finite}, {the cost of opinion changes} \cite{schawe2020collective}, agents with heterogeneous activity levels \cite{grace2023}, smooth interaction kernels (in the form of sigmoidal functions) to describe how agents influence each other \cite{brooks2022emergence}, opinion repulsion \cite{kann2023repulsive}, homophilic adaptivity of network structure \cite{kan2023adaptive}, {dynamics with ``no one left behind"} \cite{li2020bounded}, and adaptive confidence bounds \cite{grace-jerry2023}.

Another line of modeling research is the study of spreading dynamics on networks~\cite{borge2013,porter2016}. 
Research on spreading dynamics often uses ideas from percolation theory~\cite{sahimi2023,percolation2024}. The study of spreading processes on networks includes research on social dynamics \cite{lehmann2018}, disease dynamics \cite{pastor2015}, and how they affect each other \cite{bedson2021}. Much research on so-called ``social contagions'' has been inspired by the rich tradition of scholarship on biological contagions, although researchers argue about whether and how much social phenomena spread in a manner that resembles disease spread~\cite{weng2013,lerman2016,hui2018,olsson2023,cencetti2023,st-onge2024,contreras2024}. 
However, despite the differences between disease spread and the spread of information and other content, it is worthwhile to explore parallels between them. 
Whether one is considering a disease or a piece of online content, some things spread very far before dissipating and others die out very rapidly \cite{juul2022}{. {Indeed, people even say that} online content that spreads very far has ``gone viral".}

In some models of social phenomena, such as simplistic cascade models and any other models in the form of a branching process \cite{noel2012}, one can study whether the spread of content tends to magnify or contract locally as a function of time by computing branching numbers \cite{gleeson2021}, which indicate the mean number of offspring (e.g., the mean number of reposts as a function of time) of a piece of content. 
This is a simplification of empirical content spread on social media, where it is common to trace the spreading paths of posts, post boosting, and post commenting and quoting using dissemination trees \cite{gomez2010,guille2013information,goel2016structural,kozitzin2023b}. {There are also other ways to calculate local magnification and contraction in spreading processes \cite{banos2013}.}
Analogously to studies of content {spread} in social systems, when researchers study compartmental models of disease spread, it is traditional to calculate reproduction numbers to examine whether or not a disease {dies} out~\cite{brauer2019}. 
The most standard type of reproduction number is the basic reproduction number, which measures how many infections occur, on average, from one infected node in a population in which all other nodes are susceptible to infection \cite{kiss2017mathematics,brauer2019}. 
Compartmental models have also been employed in studies of opinion dynamics, and one can then calculate a basic reproduction number to examine whether or not content goes viral \cite{xu2022}. 
In the study of disease spread, branching-process theory has {also} been used in concert with maximum-likelihood estimation to estimate basic reproduction numbers from data \cite{blumberg2013}.

{In the present paper, we combine ideas from opinion models and percolation-inspired spreading models to study the mean ``infectiousness'' of content in a model that uses a bounded-confidence mechanism for content spread\footnote{A complementary approach is to modify opinion models by incorporating content~\cite{wang2024}.}. {In our content-spreading model, we} use a standard bounded-confidence mechanism to determine whether or not an agent is receptive to content. If an agent is receptive, it spreads content without modifying it, rather than compromising {its} opinion as in a standard BCM. Our {model} {takes the form of a threshold model~\cite{lehmann2018} and has an opinion-update rule that is reminiscent of bond percolation \cite{sahimi2023,dorogovtsev2008critical,newman2018networks}.
(It is also related to the independent-cascade model and its generalizations \cite{kkt2003}.)}
In {standard bond percolation,} an edge of a network is preserved with probability $p$ and is removed with probability $1 - p$. 
In our content-spreading model, the probability of ``preserving" an edge between agents $i$ and $j$ corresponds to the spread of content between those two agents. However, in contrast to classical bond-percolation models, this probability is not {fixed in our model. Instead,} it depends on the opinions of agents $i$ and $j$ and and on whether or not those two agents are adjacent to other agents that previously shared the content.} 

To analyze the structure of content {spread} in our model, we define an opinion-dynamics analogue of the basic reproduction number from disease dynamics, and we use this \emph{opinion reproduction number} and its associated 
critical value (which we call the {\emph{infodemic threshold}}) to examine when an infodemic occurs in a network. 
We use generating functions to derive an analytical expression for the opinion reproduction number on configuration-model networks in the limit of infinitely many nodes. 
We then compare this analytical result with numerical simulations. 
{We quantify the spread of content by calculating a variety of summary statistics --- the total number of shared pieces of content, the longest adoption paths, the widths of dissemination trees, and structural virality --- that were employed previously by other researchers \cite{vosoughi2018spread,juul2022}.}

Our paper proceeds as follows. In \Cref{sec:bcreview}, we give a brief introduction to bounded-confidence models of opinion dynamics. 
In \Cref{sec:process}, we describe a content-spreading process
that draws {inspiration} from percolation theory and uses a bounded-confidence mechanism for spreading. In \Cref{sec:infodemic}, we analyze the conditions that 
determine when we expect a large content-spreading cascade (i.e., an ``infodemic''). To allow us to forecast whether or not an infodemic occurs in a network, we define an opinion-dynamics analogue of the basic reproduction number from disease dynamics. In \Cref{sec:quantifyspread}, we quantify the total number of shared pieces of content on a finite-size network. In \Cref{sec:summarystats}, we compute a variety of summary statistics {to describe dissemination trees, and we} illustrate how network structure and spreading-model parameters affect these statistics in configuration-model networks. In \Cref{sec:discussion}, we conclude and discuss our results. In \Cref{david}, Mason writes a few words about David Campbell and wishes him a wonderful 80th birthday. The code to reproduce our numerical simulations is available at \url{https://github.com/hzinnbrooks/bounded-confidence-spreading-process}.


\section{Bounded-confidence models on networks} \label{sec:bcreview}

Bounded-confidence models (BCMs) \cite{lorenz2007continuous, proskurnikov2018tutorial, noorazar2020classical, noorazar2020recent} of opinion dynamics include both consensus-seeking behavior and preferences for similar views (through ``selective exposure"). BCMs have been used by many researchers to study social influence and group dynamics (including consensus, polarization, fragmentation, and other phenomena) \cite{bernardo2024}. 

We model a social network as a graph $G(V,\mathcal{E})$, where $V$ (with $\vert V \vert = N$, so $N$ is the ``size" of the network) is the set of vertices (i.e., nodes) and $\mathcal{E} \subseteq V \times V$ is the set of edges. Each node represents an agent in some population, and each edge encodes a social and/or communication tie between two agents. We suppose that $G$ is undirected {and unweighted}. Node $i$ is {\em adjacent} to node $j$ (and vice versa) if there is an edge between $i$ and $j$. We assume that $G$ is simple, so it has no self-edges or multi-edges. The graph $G$ has an associated $N \times N$ {\em adjacency matrix} $A$, where $A_{ij} = 1$ if $i$ is adjacent to $j$ and $A_{ij} = 0$ otherwise. If desired, one can {consider directed networks, weighted networks, and other} generalizations of graphs.
See Refs.~[\onlinecite{faust1994,newman2018networks,bullo2022,brooks2023}] for books and reviews about networks.

Suppose that each node $i$ has a time-dependent {opinion} $x_i(t)$, which takes continuous values on {a} domain that we call the {\em opinion space}. For example, the opinion space can be the real line $\mathbb{R}$, a closed and bounded interval $[a,b] \subset \mathbb{R}$, or a subset of a higher-dimensional space $\mathbb{R}^k$. In a BCM, agents are receptive only to other agents that are within a distance $c$ of their current opinion. {More precisely, we say that agent $i$ is {\em receptive} to agent $j$ (and vice versa) if $d(x_i(t),x_j(t)) < c$, where $d$ is a metric on the opinion space.} 
The parameter $c$ is called the {\em confidence bound}. Agents that are receptive to each other influence each other's opinions through an update rule.
It is also possible to consider BCMs with asymmetric receptiveness, but we examine only symmetric situations.


\section{A content-spreading process with a bounded-confidence mechanism} \label{sec:process}

We study a content-spreading model with a bounded-confidence mechanism for adopting an opinion and sharing content. 
However, in our model, receptive agents simply adopt the opinion that is espoused by content, rather than compromising their {opinions} as in a standard BCM.

The bounded-confidence models that {we discussed} in \Cref{sec:bcreview} describe the time-dependent opinions of the agents of a network. 
However, in real life, one {usually does not} have easy access to numerical values of people's opinions. 
One instead typically observes the content that is spread on social networks. Naturally, such content often reflects an underlying opinion or ideological belief.
There is much empirical work on quantifying the dissemination and virality of content on social-media platforms like $\mathbb{X}$ ({i.e.,} the 
platform formerly known as Twitter)~\cite{kumpel2015,muhammed2022}, 
and branching processes provide a natural model of content spread on social media. 
See, for example, Refs.~[\onlinecite{lerman2010information, guille2013information, gleeson2014,vosoughi2018spread,gleeson2021}]. 

We seek to track the spread of fixed content with a fixed 
opinion $x_0$. In other words, the spreading process has no mutation of ideas or editorializing. 
In the context of social media, one can interpret the fixed-content assumption as accounts only {sharing or reposting}
content, without any additions or changes. 
By contrast, real social-media systems have a rich ecosystem of content evolution \cite{adamic2016}.

{As we described in \Cref{sec:bcreview}, we} model a social network as {an undirected, unweighted, and simple} graph $G(V,\mathcal{E})$ with $\vert V \vert = N$ nodes. The spread of content with opinion $x_0$ begins at a source node, which we select uniformly at random from $V$ and label as node $0$. We initialize all other nodes $i \in \{1, \ldots, N - 1\}$ {of a graph $G$} with a state $x_i$ that we draw from a distribution $\phi(x)$. 
{In our analytical calculations, we assume either that there is a single source node or that the number of source nodes is sufficiently smaller than the network size $N$. We can then treat spreading processes that start from distinct nodes as independent of each other.} We require $\phi(x)$ to be Riemann integrable. 

\begin{figure*}
    \centering
    \includegraphics[width=\textwidth]{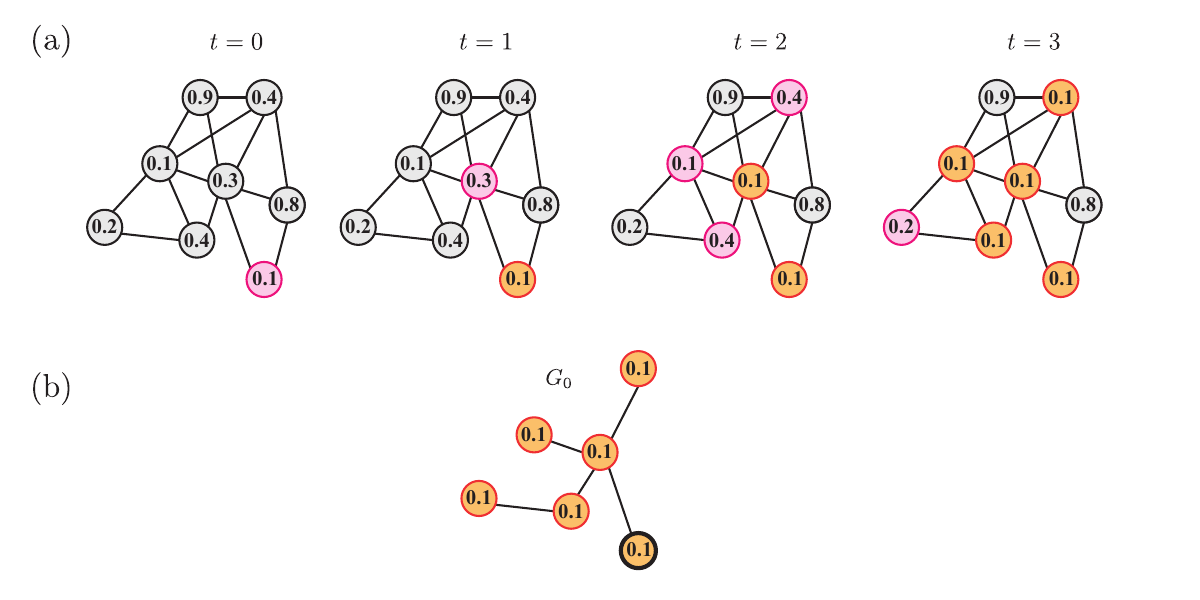}
    \caption{A schematic illustration of our content-spreading process (see \Cref{alg:update}). In panel (a), {we} show {several} consecutive update steps of \Cref{alg:update}. We initialize a graph at time $t = 0$ with one active node (in pink), which we label as node $0$. We suppose that the root node $0$ has content state $x_0 = 0.1$. We initialize all other nodes $j$ to have opinions $x_j \in (0,1)$. In this example, the receptiveness parameter is $c = 0.35$. At each time step, any neighbors $j$ of the previously active node that satisfy $\vert x_j - x_0\vert < c$ become active (i.e., turn pink). We thus add them to the dissemination tree $G_0$. In this example, \Cref{alg:update} terminates after 4 steps. In panel (b), we show the resulting dissemination tree $G_0$ in orange, with the root node outlined in black.}
    \label{fig:algorithm-schematic}
\end{figure*}

Given an opinion space $\Omega$, we refer to the opinion $x_0 \in \Omega$ of the content as the \emph{content state}. We refer to the state $x_i \in \Omega$ of node $i \in \{1, \ldots, N - 1\}$ as its \emph{opinion}. In the present paper, we focus on $\Omega = [0,1]$, although in \Cref{fig:poisson-varyx0-Gaussian} we show a situation with $\Omega = \mathbb{R}$. Let $c \in [0,1]$ denote the \emph{receptiveness parameter}, which is akin to the confidence bound of a typical BCM. In practice, in the present paper, we consider $c \in (0,1/2]$, so agents are not receptive to agents with overly different opinions.
The spread of content with opinion $x_0$ begins at node $0$. The content spreads through a percolation-inspired process \cite{sahimi2023} via an update rule that is based on a bounded-confidence mechanism (see \Cref{alg:update}). 
In \Cref{fig:algorithm-schematic}, we show a schematic of our algorithm on a small network.

\begin{algorithm}[H]
    \caption{Our algorithm for the spread of content on a network.}
    \label{alg:update}
    \begin{algorithmic}
        \State {\bf Input:} {Model parameters $c$, $x_0$, $\phi(x)$, $N$ and} 
        {the degree distribution}
        \State {\bf Output:} Dissemination tree $G_0$
        \medskip
        \State Set \texttt{active nodes} $= 0$; \texttt{next nodes} $=\{ \}$
        \While {
            \texttt{active nodes} $\neq\{\}$
            }
            \For{$i$ in \texttt{active nodes}}
	        \State {\texttt{neighbors} $=\{ j \ \vert \ A_{ij} = 1 \}$}
	        \For{$j$ in \texttt{neighbors}}
	        \vspace{-4mm}
	        \State {\[ x_{j} = \begin{cases}
	x_i  \text{
	{(add $i$ to \texttt{next nodes})}}\,,
	& \text{if } \vert x_j - x_0 \vert < c  \\
	x_{j}\,, & \text{otherwise}
	\end{cases}  \]}
	 \vspace{-2mm}
	        \EndFor
	        \For{$k$ in \texttt{next nodes}}
	        \State {$A_{ik} = 0$ for all $i$}
	        \EndFor
            \EndFor
        \State \texttt{active nodes} $=$ \texttt{next nodes} 
        \State Add \texttt{next nodes} to $G_0$
        \State \texttt{next nodes} $=\{\}$
        \EndWhile
    \end{algorithmic}
\end{algorithm}

To study content spread on a network, we construct a {\em dissemination tree} $G_0$ with root node $0$
 to track the spreading process~\cite{goel2016structural,oh2018complex}.
Content spreads as follows. Suppose that content state $x_0$ starts at node $0$.  
We look at the set $\mathcal{N}_i$ of neighbors $j$ of node $i$ and select all {nodes} $j \in \mathcal{N}_i$ {with an opinion $x_j$ that satisfies} $\vert x_j - x_0 \vert < c$. 
These neighbors of $i$ spread the content (i.e., they {``activate"} and change their {opinions} $x_j$ {to} the content state $x_0$) {because the} distance between {the} opinion $x_j$ and the content state $x_0$ is sufficiently small. 
We represent this activation by adding these nodes to the dissemination tree $G_0$ as child nodes of node $i$. 
We then continue this process for each child node of the current dissemination tree until there are no available neighbors that are left to activate. 
In other words, we continue this process until either all nodes are activated or until {no active nodes have any remaining receptive neighbors.}

Using \Cref{alg:update}, it is guaranteed that $G_0$ is a tree because a node activates the first time that it encounters the content of one of its neighbors and can only {activate once.} When two {active} nodes share a receptive neighbor, we choose the {active} node with the smaller index as the parent node {of} the newly activated neighbor. Modifications of this algorithm can result in a directed acyclic graph instead of a tree. It is worthwhile to examine such modifications in future work ({e.g.,} if considering the effects of competing social contagions \cite{min2018competing}).


\section{Forecasting an ``infodemic'' on a large network} \label{sec:infodemic}

We now examine content spread on a network with infinitely many nodes. 
This situation approximates content spread on networks with many nodes (i.e., {large} $N$), particularly in the early stages of content spread. 

We use analysis {that was developed for models of infectious-disease spread on networks \cite{newman2001random, newman2002spread, noel2009time}.} We use generating functions to 
{obtain an expression for} the expected number of neighbors that spread the source node's content. We refer to this expected number of neighbors as the {\em opinion reproduction number} $R$. 
The opinion reproduction number, whose critical value $R = 1$ is its associated \emph{infodemic threshold}, is akin to the basic reproduction number $R_0$ in models of disease spread \cite{kiss2017mathematics,brauer2019}.


\subsection{Analysis} \label{sec:infodemic-analysis}

Suppose that content spreads on a configuration-model network~\cite{fosdick2018} with degree distribution $p_k$. We generate each configuration-model network {by uniformly randomly matching ``stubs" (i.e., ends of edges) to each other.} {We remove all self-edges and multi-edges} {after matching stubs, so different networks can have slightly different degree sequences.} Let $q_k$ denote the associated excess degree distribution. 
The generating function $g_0$ of the degree distribution and the generating function $g_1$ of the excess degree distribution $q_k$ are
\begin{align}
	\label{eqn:g0}
	g_0(z) = \sum_{k = 0}^\infty p_k z^k \,, \\
	\label{eqn:g1}
	g_1(z) = \sum_{k = 0}^\infty q_k z^k\,.
\end{align}

Given a node with $k$ neighbors, we want to {determine} the probability that $l$ of these $k$ neighbors spread a piece of content.
To {do} this, we need to calculate the probability that content is spread along one edge. 
This single-edge transmission probability depends {on} the content state $x_0$, the receptiveness parameter $c$, and the distribution of the initial opinions $x_i$ (with $i \in \{1, \ldots, N\}$). 

Suppose that we draw the initial opinions $x_i$ from a distribution with probability density function $\phi(x)$.
The probability of content spread (i.e., transmission) along an edge depends on the probability of drawing a value $x_i$ that lies within {a} distance $c$ of $x_0$. 
The single-edge transmission probability is
\begin{equation}
	s(x_0,c) = \int_{x_0 - c}^{x_0 + c}\phi(x) \, dx \,.
	\label{eqn:probshare}
\end{equation}

One common choice of $\phi(x)$ is to draw the initial node opinions from the uniform distribution on the interval $(0,1)$.
The probability of drawing an opinion $x_i$ within {a} distance $c$ of $x_0$ is then $2c$, unless $x_0$ is within $c$ of the boundaries of $(0,1)$.
This yields
\begin{equation}
	s(x_0,c) = \begin{cases}
	c + x_0 \,, & x_0 \leq c \\
	2c \,, & x_0 \in (c,1 - c]  \\
	1 + c - x_0 \,, & x_0 > 1 - c\,.
	\end{cases}
\label{eqn:unifprob}
\end{equation}
Another common choice of  $\phi(x)$ is to draw the initial node opinions from a Gaussian distribution with mean $\mu$ and standard deviation $\sigma${. (In this case, the opinion space is $\mathbb{R}$.)} This yields
\begin{equation}
	s(x_0,c) = \frac{1}{2}\left[ \mathrm{erf}\left(\frac{x_0 + c - \mu}{\sigma\sqrt{2}}\right) - \mathrm{erf}\left(\frac{x_0 - c-\mu}{\sigma\sqrt{2}}\right)\right] \,,
\end{equation}
where $\mathrm{erf(x)} = \frac{2}{\sqrt{\pi}}\int_0^xe^{-t^2} \, dt$ denotes the error function. 

Using the single-edge transmission probability [see \Cref{eqn:probshare}], we see that the probability that $l$ of $k$ neighbors spread the content is
\begin{equation}
	p(l \vert k) = \binom{k}{l}\left(s(x_0,c)\right)^l\left(1 - s(x_0,c)\right)^{k - l} \,.
	\label{eqn:plk}
\end{equation}
Therefore, the probability generating function of the distribution of the number of content shares that originate from the source node $0$ is
\begin{widetext}
\begin{align}
	\sum_{l=0}^\infty \sum_{k=l}^\infty p_k p(l \vert k) z^l &= \sum_{l=0}^\infty \sum_{k=l}^\infty p_k \binom{k}{l}\left(s(x_0,c)\right)^l\left(1-s(x_0,c)\right)^{k-l} z^l \nonumber \\
&=\sum_{k=0}^\infty p_k \sum_{l=0}^k \binom{k}{l}\left(s(x_0,c)\right)^l\left(1-s(x_0,c)\right)^{k - l} z^l  \nonumber \\
&= \sum_{k=0}^\infty p_k\left(1+(z-1)s(x_0,c)\right)^k \nonumber\\
	&= g_0\left(1+(z-1)s(x_0,c)\right) \,.
\label{eqn:pgf_0}
\end{align}
\end{widetext}

To {obtain} the probability generating function for the number of second neighbors (i.e., the neighbors of neighbors) of a source node in a dissemination tree, we use a similar argument, but now we use the excess degree distribution instead of the degree distribution. {This yields the probability generating function}
\begin{equation}
	g_1\left(1 + (z - 1)s(x_0,c)\right) \,.
	\label{eqn:pgf_1}
\end{equation}

If a network $G$ is infinite (or at least large enough so that we can neglect finite-size effects), nodes cannot {activate} more than once, and the content spread yields a 
directed acyclic graph (i.e., a dissemination tree). {Consequently,} $g_a\left(1 + (z - 1)s(x_0,c)\right) = g_1\left(1 + (z - 1)s(x_0,c)\right)$ is the probability generating function for the number of nodes that spread a piece of content that they received directly from a single node at level $a$ of a dissemination tree. Importantly, we are relying on the assumption that the number of times that a piece of content is spread is much smaller than the total number of nodes of a network.

{In the early stages of} content dissemination, the expected number of ``grandchildren" (i.e., second neighbors) in a dissemination tree that starts from a node that spreads the 
content at level $a$ is the opinion reproduction number
\begin{align}
	R = \frac{\mathrm{d}}{\mathrm{d}z}g_a\left(1 + (z - 1)s(x_0,c)\right) \bigg \vert_{z = 1} &= s(x_0,c)g_1'(1) \nonumber \\
	\label{eqn:second_messages}
		&= s(x_0,c)\frac{g_0''(1)}{g_0'(1)} \,,
\end{align}
where the prime $'$ denotes differentiation with respect to $z$. {The right-hand side of \Cref{eqn:second_messages} relates the expected number of grandchildren to the expected number of children.} When the number of content shares increases from one level to the next, the opinion reproduction number $R$ (which is analogous to the basic reproduction number $R_0$ in models of infectious-disease spread \cite{kiss2017mathematics}) is larger than $1$. 
Similarly, a decrease in {the} number of content shares from one level to the next corresponds to $R < 1$.
The critical value $R = 1$ is the \emph{infodemic threshold}. This calculation is reminiscent of branching-processes approaches that examine whether or not content spread tends to locally magnify or contract with time \cite{gleeson2021}.
{Given an opinion reproduction number $R$, we approximate the expected number of content shares up to and including time step $T$ as $n_0\sum_{t = 0}^T R^t$, where $n_0$ is the number of source nodes and we treat content that spreads from different source nodes as independent. Given our assumptions, we expect our approximation to be reasonable in the ﬁrst few generations of a dissemination tree. These initial generations describe the early stages of content spread.}

We now calculate $R$ explicitly for an example in which we draw the initial node opinions uniformly at random from the interval $(0,1)$. 
{In a configuration-model network with a Poisson degree distribution with mean $\lambda$, the opinion reproduction number is 
\begin{equation} \label{pois}
	R = s(x_0,c)\frac{\lambda^2}{\lambda} = s(x_0,c)\lambda
\end{equation}	 
because $g_0(z) = e^{-\lambda(1 - z)}$. 
{Substituting \Cref{eqn:unifprob} into \Cref{pois} allows us to calculate the opinion reproduction number in terms of the content state $x_0$, the receptiveness parameter $c$, and $\lambda$. 
At the infodemic threshold $R = 1$, the critical value $c^*$ of the receptiveness parameter that} determines whether content spread increases locally or decreases locally is
\begin{equation*} 
	c^* = 	\begin{cases} \frac{1}{\lambda} + x_0 \,, & x_0\leq c^* \\
 		\frac{1}{2\lambda}\,, & c^* < x_0 \leq 1 - c^*  \\
		\frac{1}{\lambda} - 1 + x_0\,, & x_0 > 1 - c^* \,.
			\end{cases} 
\end{equation*}
When the receptiveness parameter $c > c^*$, the expected number of content shares increases from one level to the next in the early stages of content spread, so we expect the content to take hold in a network. Unsurprisingly, the infodemic threshold $R = 1$ is exceeded most easily for larger values of the receptiveness parameter $c$ and networks with larger expected mean degree $\lambda$.} The content state $x_0$ has less impact than the receptiveness parameter $c$ and {the} {expected mean} degree, except when the content state is near $0$ or $1$.


\subsection{Simulations}

\begin{figure*}[htbp]
  \centering
  \includegraphics[width=\textwidth]{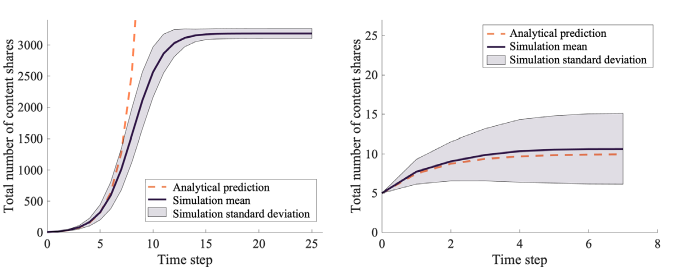}
  \caption{{A comparison of numerical simulations of our content-spreading model with our analytical predictions in (a) an infodemic situation (when the receptiveness parameter  is $c = 0.2$, which implies that $R > 1$) and (b) a situation without an infodemic (when the receptiveness parameter is $c = 0.05$, which implies that $R < 1$).
  For each numerical simulation, we generate a 10000-node configuration-model network with a degree sequence that we determine using a Poisson distribution with mean $\lambda = 5$. We draw the initial node opinions $x_i$ uniformly at random from 
  $(0,1)$, and we set the content state to $x_0 = 0.5$. In each realization, we choose $5$ source nodes uniformly at random to seed with content state $x_0$. {In our analytical calculations, we assume that the content spread from each source node is independent, as the number of source nodes is much smaller than the total number of nodes.}
  The solid purple curve gives the total number of content shares {(averaged over $100$ realizations) as a function of time.} 
  In each realization, we draw a new degree sequence and a new set of node opinions.
  The light purple shaded region indicates the standard deviation for these $100$ realizations.
  The dashed orange curve gives our analytical approximation.}
  }
  \label{fig:poissondegreedist} 
\end{figure*}

We now compare our analytical results from \Cref{sec:infodemic-analysis} to numerical simulations of our content-spreading model. 
We simulate the model on 10000-node configuration-model networks with degree sequences that we determine using a Poisson distribution with mean $\lambda = 5$. 
We draw the initial opinions uniformly at random from the interval $(0,1)$. In {\Cref{fig:poissondegreedist}(a)}, we show a scenario in which our analysis 
indicates that we are above the infodemic threshold (i.e, $R > 1$). In {\Cref{fig:poissondegreedist}(b)}, we show a scenario in which our analysis indicates that we are below the infodemic threshold (i.e., $R < 1$).
In this computation, the mean total number of content shares is very small. (It is about $10$.)

\begin{figure}[htbp]
  \centering
  \includegraphics[width=0.48\textwidth]{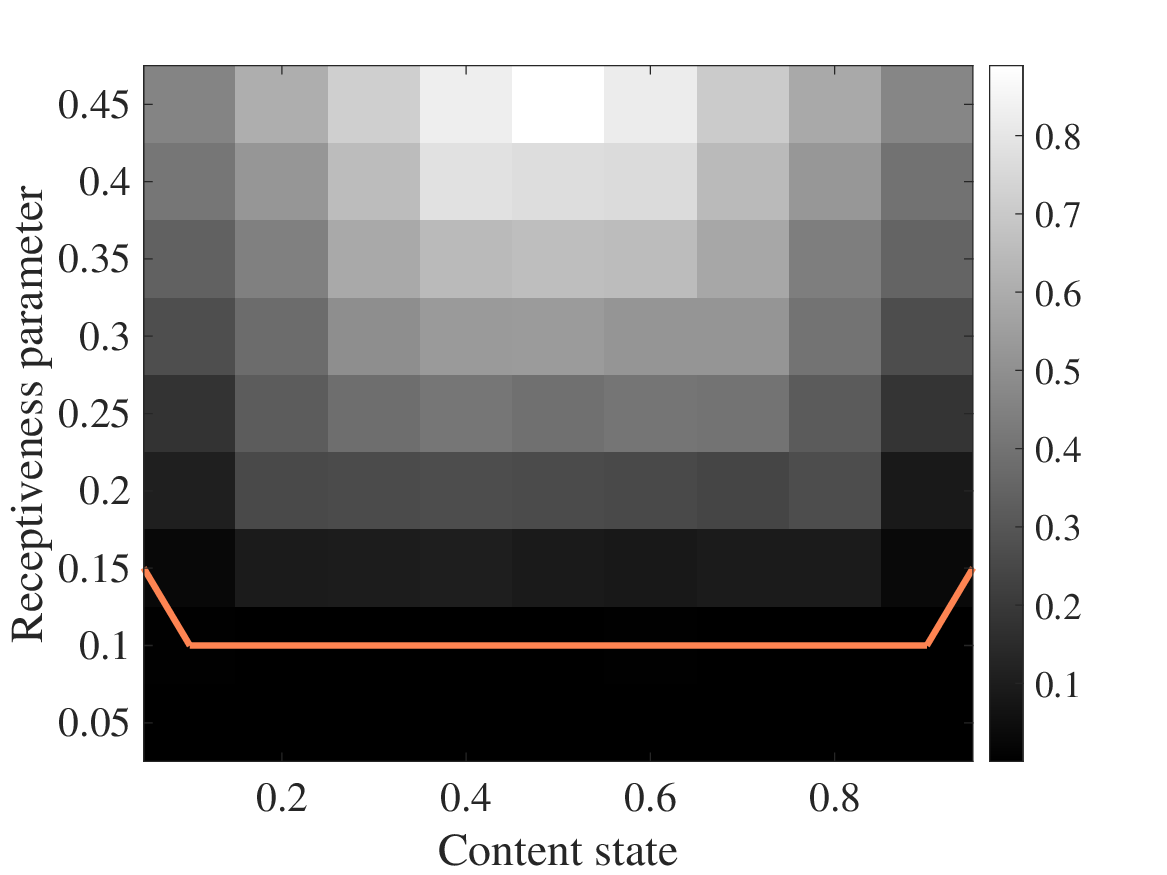}
  \caption{The $(x_0,c)$ phase diagram for 10000-node configuration-model networks with degree sequences that we generate using a Poisson distribution with mean $\lambda = 5$. {We vary both the content state $x_0$ 
  and the receptiveness parameter $c$.} The shading of each square indicates the mean, across 100 trials, of the total number of content shares as a proportion of the total number of nodes, with lighter shades indicating that more nodes share the content.
  In each realization, we choose $5$ source nodes uniformly at random to seed with content state $x_0$. In each realization, we draw a new degree sequence and a new set of node opinions.
  The orange solid curve {shows the critical receptiveness value $c^*$ for each $x_0$.}}
  \label{fig:poissondegreedist-pp} 
\end{figure}

In \Cref{fig:poissondegreedist-pp}, we show a phase diagram that summarizes when our content-spreading model {experiences} an infodemic for different values of the content state $x_0$ and receptiveness parameter $c$. 
{In orange, we show the critical value $c^*$ that we obtain analytically using the infodemic threshold $R = 1$.}

\section{Quantifying content spread} \label{sec:quantifyspread}

Now that we have predicted the onset of infodemics in our content-spreading model on configuration-model networks, we measure the sizes (i.e., the numbers of nodes) in the resulting dissemination trees. 
The size of a dissemination tree equals the total number of content shares.
Studying dissemination trees allows us to explore the long-term behavior of our content-spreading model. These calculations thereby complement our examination of early-stage spreading in \Cref{sec:infodemic}.


\subsection{Analysis of the total number of content shares}

Given the number of nodes and the degree distribution of a configuration-model graph $G$, {a} content state $x_0$, {a} receptiveness parameter $c$, and {a} distribution $\phi(x)$ of opinions, we again use generating functions (as in \Cref{sec:infodemic-analysis}) to estimate the total number of content shares (i.e., the size) of a dissemination tree $G_0$. 
This analysis treats content spread as a percolation process. See Ref.~[\onlinecite{newman2018networks}] for a detailed description of this approach.

{Let the ``spreading set" $\mathcal{S}$ of a network denote the set of nodes that have shared a piece of content. The spreading set $\mathcal{S}$ has $SN$ elements (i.e., there are a total of $SN$ content shares), where $S$ is the fraction of a network's nodes that have shared that content.} Node opinions are independent of connectivity in our content-spreading model, so the probability that a node is in $\mathcal{S}$ equals the probability that it is receptive to the content multiplied by the probability that it belongs to the connected component of a source node. {In this {calculation, we assume that there is a single source node. Therefore, $SN$} is the size of the connected component of one source node.}
Let $u$ be the mean probability of not being connected to this component via a particular neighbor. We then have
\begin{align}
	S &= \left(\int_{x_0 - c}^{x_0 + c}\phi(x) \, dx\right)\left(1 - \sum_{k=0}^\infty p_ku^k\right) \nonumber \\
		&= s(x_0,c)\left(1 - g_0(u)\right) \,, \label{eqn:s-eqn}
\end{align}
where we recall that $p_k$ is the probability {that} a node has degree $k$ and $g_0(u)$ is the generating function of the degree distribution of the {graph}.

To calculate $S$, we need to determine $u$. A particular node $i$ {is not connected} to the spreading set $\mathcal{S}$ via a particular node $j$ {either} because it is not receptive to the content or because it is not in the same component as the content. 
{In the latter case, node $i$ is} isolated from the content spread because of non-receptive nodes. For a node $j$ with $k$ edges aside from the one to node $i$, {the probability that one of these scenarios occurs is} $1 - s(x_0,c) + s(x_0,c)u^k$.
Therefore, 
\begin{align}
\label{eqn:u-eqn}
	u &= \sum_{k=0}^\infty q_k\left(1 - s(x_0,c) + s(x_0,c)u^k\right) \nonumber \\
		&= 1 - s(x_0,c) + s(x_0,c)g_1(u) \,, 
\end{align}
where $q_k$ again denotes the probability of having $k$ neighbors other than the spreading edge {(i.e., the excess degree is $k$)} and $g_1(u)$ again denotes the generating function of the excess degree distribution. By inspection, $u = 1$ is always a solution; this solution entails not having an infodemic. We are interested in whether or not there are also solutions $u \in (0,1)$. 

In some special cases, it is possible to solve for $u$ analytically, but it is typically difficult 
(or even impossible) to do so even when $g_1(u)$ has a simple closed-form expression. 
However, {one can find a} root $u^* \in (0,1)$ of \Cref{eqn:u-eqn} numerically {either using an explicit expression or using an approximation} of the generating function of the excess degree distribution~\cite{newman2018networks}. One then substitutes $u^*$ into \Cref{eqn:s-eqn} to obtain an approximation of $S$. The solution {$u^* \in (0,1)$} yields an approximation of $S$ when there is an infodemic.

\begin{figure}[t]
    \centering
    \includegraphics[width=0.48\textwidth]{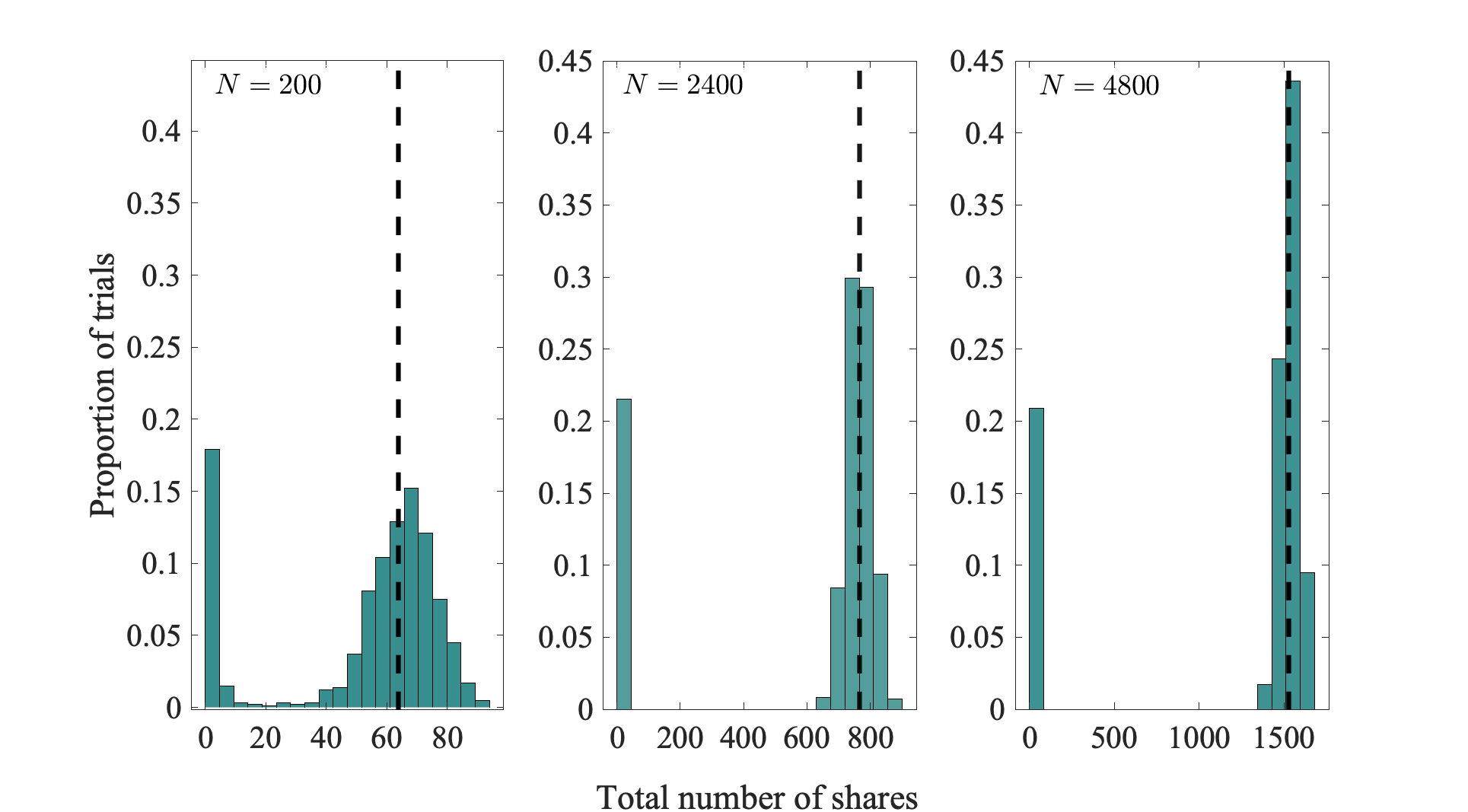}
    \caption{Histograms of the total number of content shares {in our content-spreading model} across 1000 trials for 200-node, 2400-node, and 4800-node configuration-model networks. 
    The dashed vertical {lines indicate the numbers} of content shares (i.e., the dissemination-tree sizes) that we predict from our analysis.
    In each realization, we generate an $N$-node configuration-model network with a degree sequence that we obtain from a Poisson distribution with mean $\lambda = 5$. 
    We draw the node opinions $x_i$ uniformly at random from $(0,1)$, and we set the content state {to} $x_0 = 0.5$
    and the receptiveness parameter {to} $c = 0.2$. In each realization, we draw a new degree sequence and a new set of node opinions.
    }
    \label{fig:analytic-shares}
  \end{figure}

In \Cref{fig:analytic-shares}, we show an example {in which} we compare our approximation of the total number of content shares ($SN$) to numerical simulations in configuration-model networks of three different sizes. For each size (200, 2400, and 4800 nodes), we generate 1000 configuration-model networks with degree sequences from a Poisson distribution with mean $\lambda = 5$. We plot histograms of the total number of content shares, and we observe good agreement between our analytical and numerical results.

\section{Summary statistics for dissemination trees} \label{sec:summarystats}

\begin{figure}[htbp]
  \centering
  \includegraphics[width=0.48\textwidth]{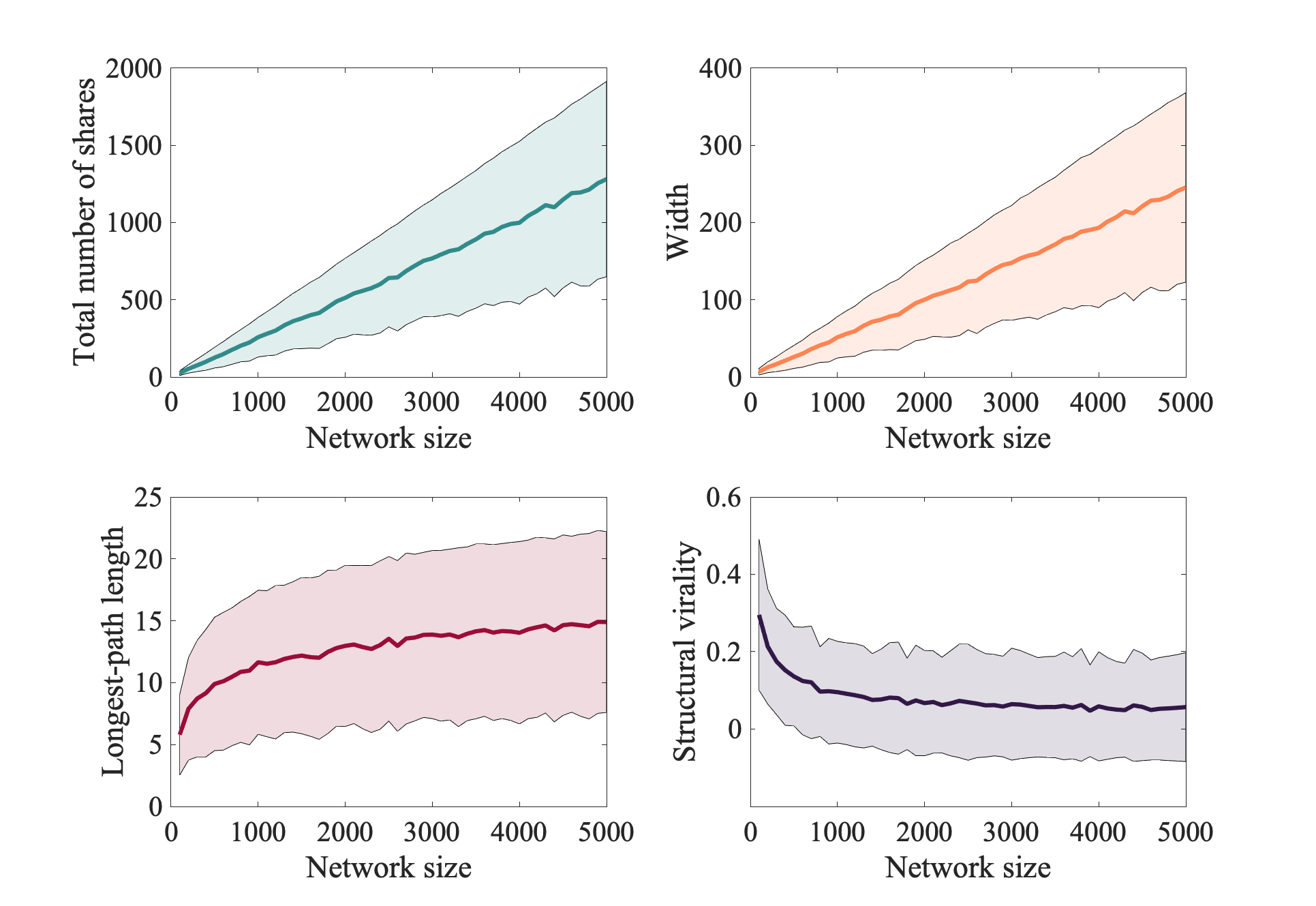}
  \caption{The effect of varying the network size $N$ on the total number of content shares, the width, the longest-path length, and the structural virality of dissemination trees of our content-spreading model on configuration-model networks. The solid curves give means across $1000$ realizations, and the shaded regions give the standard deviations. 
  In each realization, we generate an $N$-node configuration-model network with a degree sequence from a Poisson distribution with mean $\lambda = 5$.  We vary $N$ from $100$ to $5000$ in increments of $100$. Each realization has different initial node opinions, which we draw uniformly at random from $(0,1)$. 
  {The content state is $x_0 = 0.5$ and the receptiveness parameter is $c = 0.2$. In each realization, we draw a new degree sequence and a new set of node opinions.}
    Both the total number of content shares (i.e., the dissemination-tree size) and the width grow linearly
  with $N$. The longest-path length grows quickly at first as we increase $N$, and then it grows much more slowly with $N$.
  The structural virality appears to saturate at a constant value for sufficiently large $N$.
  }
  \label{fig:poisson-varyN} 
\end{figure}

\begin{figure}[htbp]
  \centering
  \includegraphics[width=0.48\textwidth]{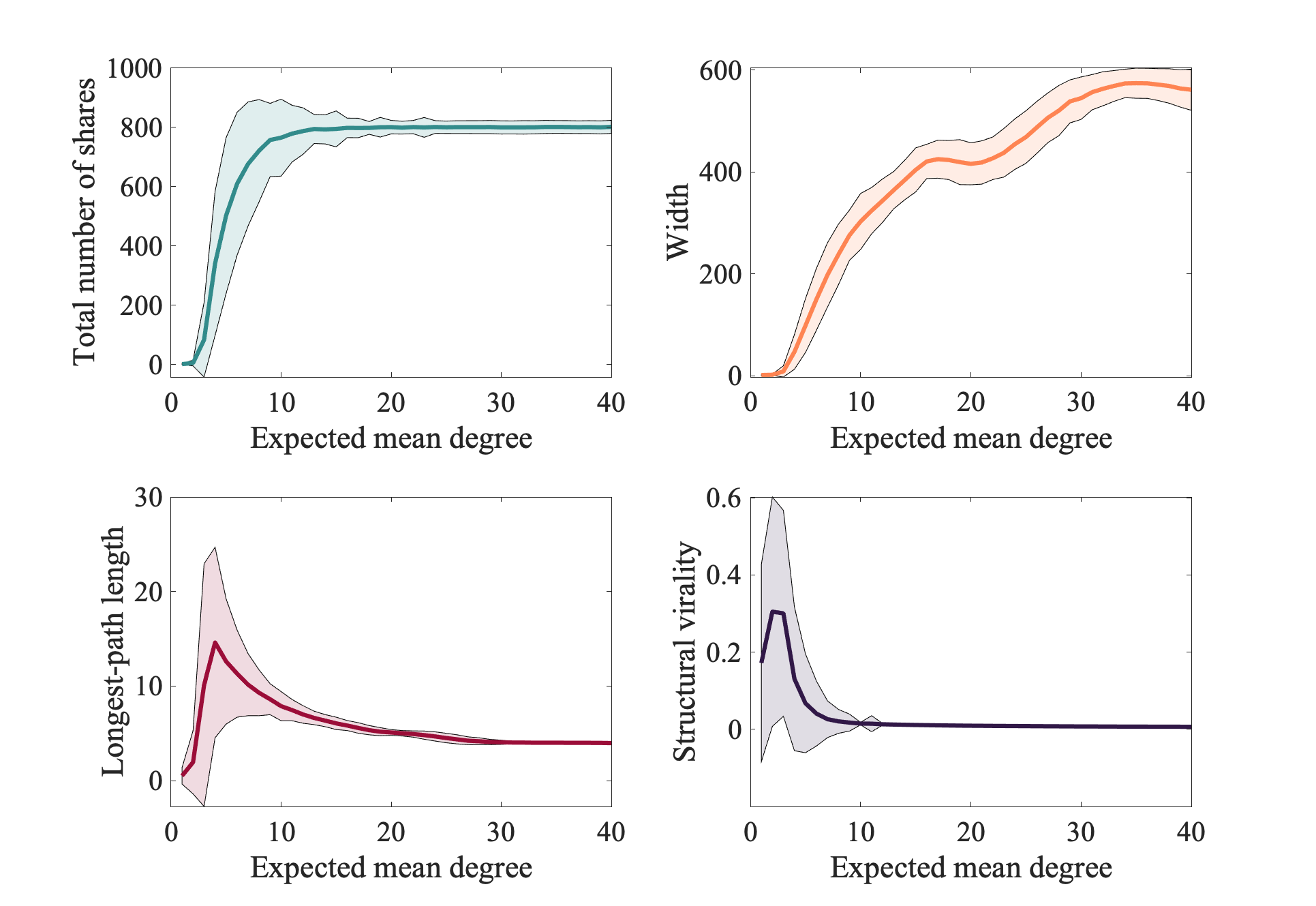}
  \caption{The effect of varying the expected mean degree $\lambda$ of a network on the total number of content shares, the width, the longest-path length, and the structural virality of dissemination trees of our content-spreading model on configuration-model networks. The solid curves give {means} across $1000$ realizations, and the shaded regions give the standard deviations. In each realization, we generate a 2000-node configuration-model network with a degree sequence from a Poisson distribution with mean $\lambda$. We vary $\lambda$ from $1$ to $40$ in increments of $1$. Each realization has different initial node opinions, which we draw uniformly at random from $(0,1)$. 
  {The content state is $x_0 = 0.5$ and the receptiveness parameter is $c = 0.2$. In each realization, we draw a new degree sequence and a new set of node opinions.}
 The total number of content shares grows initially and then saturates at about $800$. 
  The width also tends to increase with $\lambda$, with a possible plateau in a small interval near $\lambda = 20$.
  The longest-path length and structural virality increase at first and then decrease, eventually leveling off at a constant value for sufficiently large expected mean degrees. In our simulations, the longest-path length has a maximum at about $\lambda = 4$ and {the} structural virality has a maximum at about $\lambda = 3$.}
  \label{fig:poisson-varyk} 
\end{figure}

\begin{figure}[htbp]
  \centering
  \includegraphics[width=0.48\textwidth]{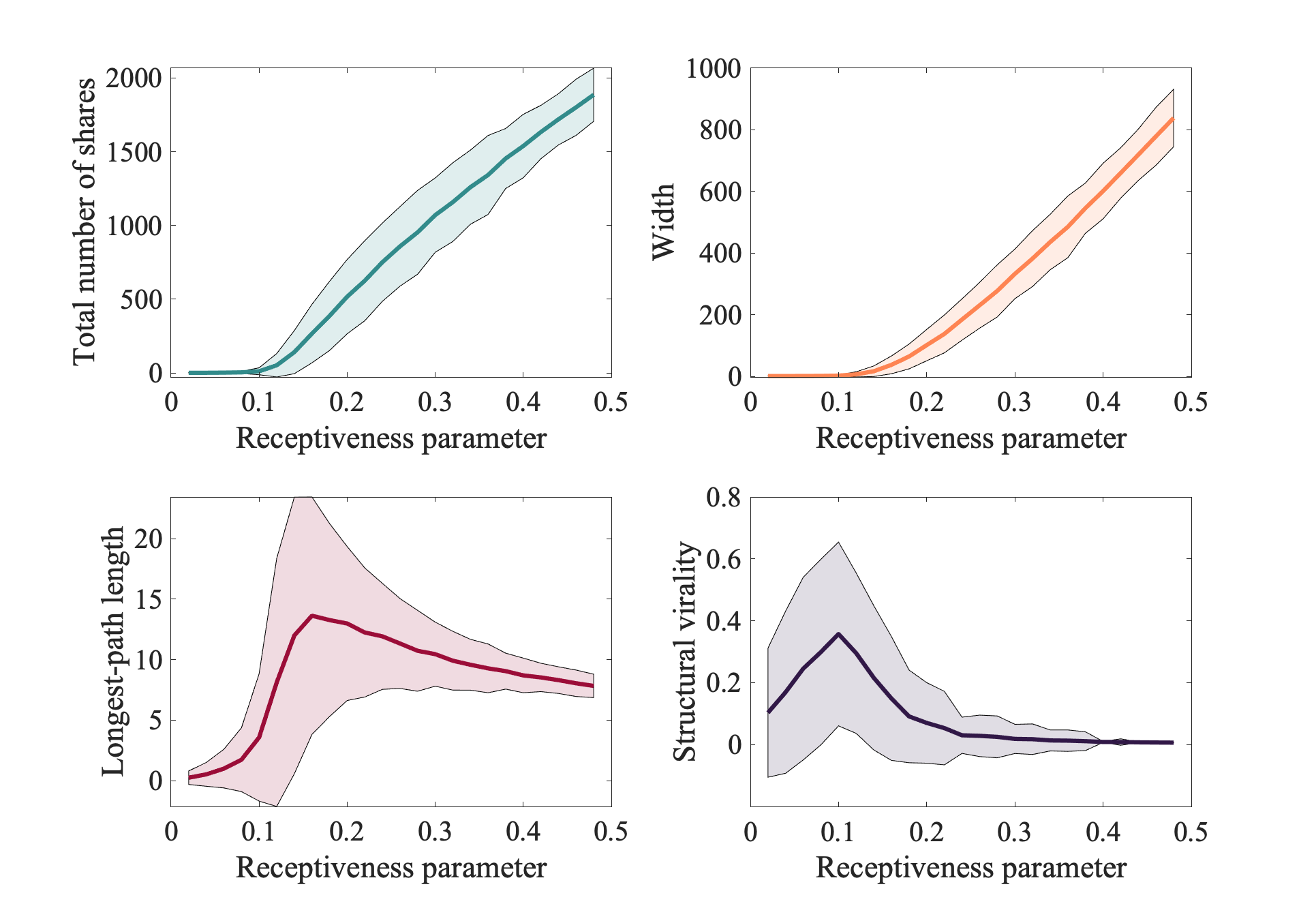}
  \caption{The effect of varying the receptiveness parameter $c$ on the total number of content shares, the width, the longest-path length, and the structural virality of dissemination trees of our content-spreading model on configuration-model networks. {The solid curves give means across $1000$ realizations, and the shaded regions give the standard deviations. In each realization, we generate a 2000-node configuration model network with a degree sequence from a Poisson distribution with mean $\lambda = 5$.} Each realization has different initial node opinions, which we draw uniformly at random from $(0,1)$. 
    The content state is $x_0 = 0.5$. We vary $c$ from $0.02$ to $0.48$ in increments of $0.02$.
    In each realization, we draw a new degree sequence and a new set of node opinions.
The total number of content shares (i.e., the dissemination-tree size) and the width are very small until about $c = 0.1$, and then they grow with $c$. Both growth rates seem roughly linear {after some earlier slow growth}, although the curve for the total number of content shares seems to be concave down. The longest-path length increases initially with $c$ before reaching
  a maximum and then decaying. The structural virality also appears to increase initially with $c$ before reaching a maximum and then decaying to a constant value.
  The standard deviations of the longest-path length and the structural virality are large.
  }
  \label{fig:poisson-varyc} 
\end{figure}

\begin{figure}[htbp]
  \centering
  \includegraphics[width=0.48\textwidth]{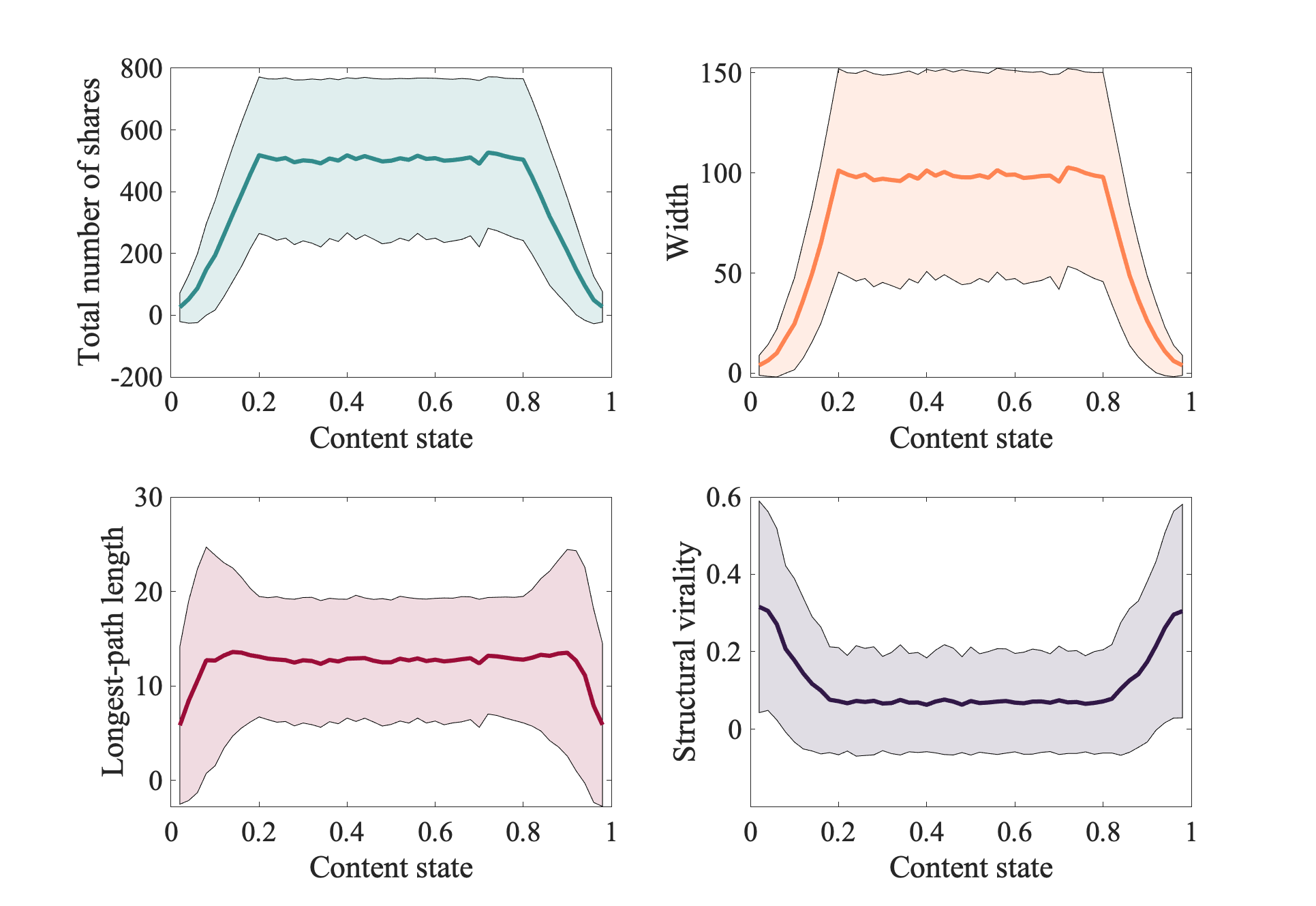}
  \caption{The effect of varying the content state $x_0$ on the total number of content shares, the width, the longest-path length, and the structural virality of dissemination trees of our content-spreading model on configuration-model networks with initial node opinions from a uniform distribution. {The solid curves show {means} across $1000$ realizations, and the shaded regions give the standard deviations. In each realization, we generate a 2000-node configuration-model network with a degree sequence from a Poisson distribution with mean $\lambda = 5$. Each realization has different initial node opinions, which we draw uniformly at random from $(0,1)$. 
   {The receptiveness parameter is $c = 0.2$.} We vary $x_0$ from $0.02$ to $0.98$ in increments of $0.02$. In each realization, we draw a new degree sequence and a new set of node opinions.}
  When $x_0 \in (0.2, 0.8)$, {the total number of content shares, the width, the longest-path length, and the structural virality are all roughly constant.}
  This is not surprising because we draw initial node opinions uniformly at random, so the transmission probability for content to spread along an edge depends only
   on $c$ in this interval (see \Cref{eqn:unifprob}). The symmetry of these summary statistics with respect to $x_0$ is also clear from \Cref{eqn:unifprob}. 
   Content states that are closer to the boundary {of opinion space} produce dissemination trees with fewer total content shares, smaller widths, and shorter longest paths. However, the structural virality of such ``extreme" content states is slightly larger.
  }
  \label{fig:poisson-varyx0} 
\end{figure}

\begin{figure}[htbp]
  \centering
  \includegraphics[width=0.48\textwidth]{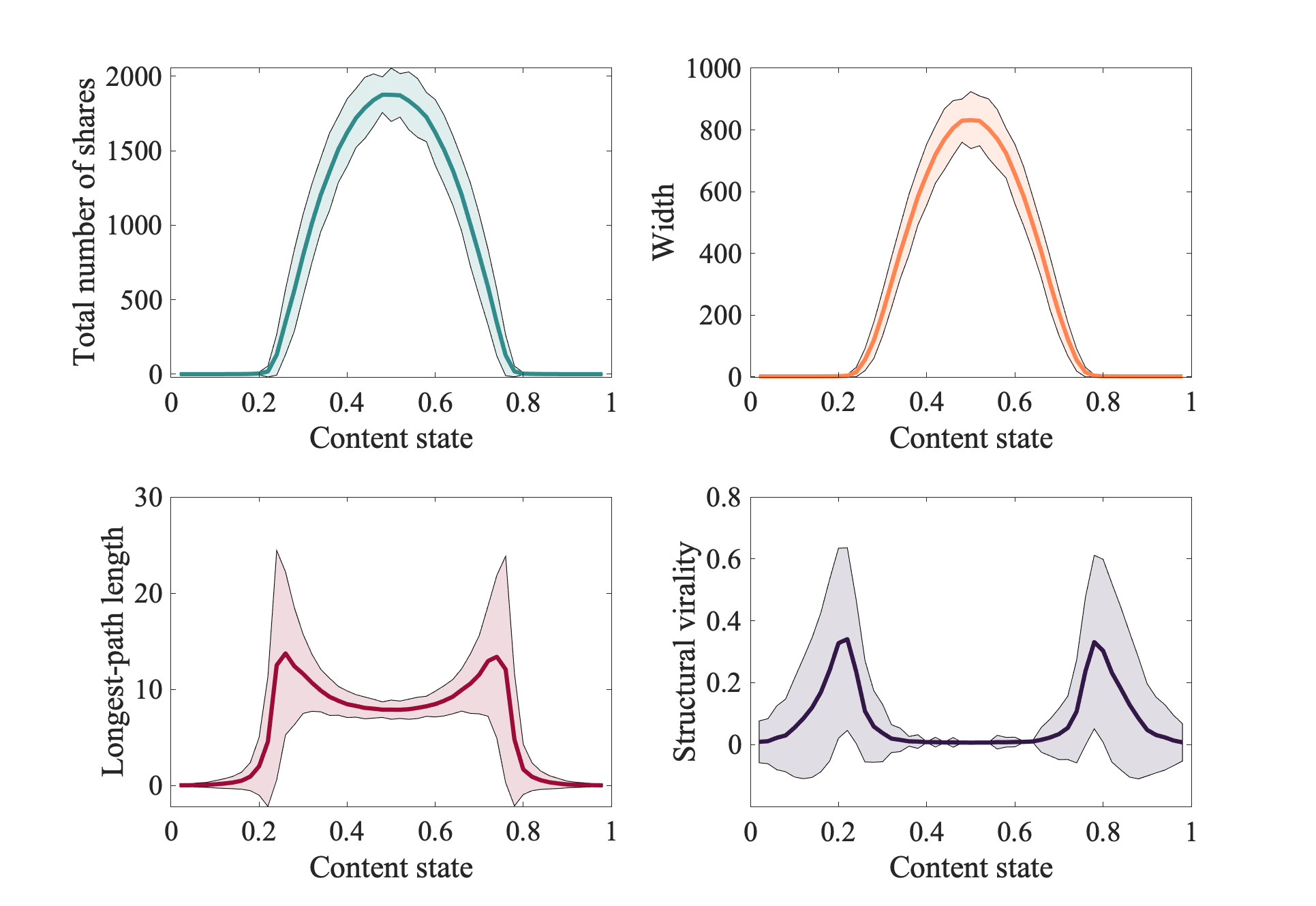}
  \caption{The effect of varying the content state $x_0$ on the total number of content shares, the width, the longest-path length, and the structural virality of dissemination trees of our content-spreading model on configuration-model networks with initial node opinions that we draw from a Gaussian distribution with mean $0.5$ and standard deviation $0.1$. {In this figure, the opinion space is
  $\mathbb{R}$. This opinion space and the initial opinion distribution differ from the ones in \Cref{fig:poisson-varyx0}.}
  The solid curves give means across $1000$ realizations, and the shaded regions give the standard deviations. In each realization, we generate a 2000-node configuration-model network with a degree sequence from a Poisson distribution with mean $\lambda = 5$. {The receptiveness parameter is $c = 0.2$.}
  We vary the content state $x_0$ from $0.02$ to $0.98$ in increments of $0.02$.
  In each realization, we draw a new degree sequence and a new set of node opinions.
   }
  \label{fig:poisson-varyx0-Gaussian} 
\end{figure}

In this section, we perform numerical experiments and study the output of our bounded-confidence content-spreading model on configuration-model
networks. We quantify its features by computing four summary statistics of the dissemination trees that we obtain in our simulations.
\begin{itemize}
	\item {\bf Total number of content shares.} One way to measure the effectiveness of content spread is to count the total number of nodes that spread (i.e., ``adopt'') the content. This quantity, which we studied in \Cref{sec:quantifyspread}, is equal to the number of nodes in the dissemination tree $G_0$.
	\item {\bf Length of a longest adoption path.} An adoption path is a 
	path in a dissemination tree from a source node to another node in the tree~\cite{oh2018complex}. 
    We measure the ``depth" of content spread by calculating the length of a longest adoption path.
    	\item {\bf Width.} One can arrange a dissemination tree $G_0$ with the source node at the top (i.e., at level $0$), nodes with adoption-path length $1$ at level $1$, nodes with adoption-path length $2$ at level $2$, and so on. 
	The width of a dissemination tree is the largest number of nodes that adopt the content in a level. Mathematically, the width is
\begin{equation*}
	\max_{{b} \in \{1, \ldots, l_*\}} \text{(number of nodes with {adoption-path} length} {\,\, b})
	 \,, 
\end{equation*}
where $l_{*}$ is the length of {a} longest adoption path.
	\item {\bf Structural virality.} Structural virality (i.e., the Wiener index) was used by Goel et al.~\cite{goel2016structural} to measure the viral nature of content spread. 
	The structural virality $v$ is the mean shortest-path length between nodes
	in {a} dissemination tree $G_0$. Mathematically, the structural virality is
\begin{equation*} 
	v = \frac{1}{n(n -1)}\sum_{i = 1}^N\sum_{j = 1}^N d_{ij} \,, 
\end{equation*}
where $d_{ij}$ is the length of a shortest path between nodes $i$ and $j$.
\end{itemize}

One natural question to ask is how the number of nodes of a network affects the spread of content on that network.
We study this question by increasing the network size $N$ for fixed expected mean degree $\lambda$ (see \Cref{fig:poisson-varyN}).
With this construction, the expected number of edges remains constant as we increase $N$.
As implied by our analysis in \Cref{sec:quantifyspread}, the total number of content shares grows linearly with $N$ (with a slope indicated by \Cref{eqn:s-eqn}). 
The dissemination-tree width also appears to grow linearly with $N$. 
The longest-path length and structural virality change more dramatically for small $N$ than for large $N$. The former appears to grow very slowly for large $N$, and the latter appears to eventually saturate at a constant value.

In \Cref{fig:poisson-varyk}, we examine the impact of increasing the expected mean degree $\lambda$ on the dissemination-tree statistics. The total number of content shares increases with $\lambda$ before saturating at a constant value. The width also tends to increase with $\lambda$, with a possible plateau in a small interval near $\lambda = 20$.
The longest-path length and structural virality have maxima for small values of $\lambda$.
When $\lambda$ is large, it is seemingly possible {for} the content to be adopted by all receptive and connected nodes in just a few steps, yielding a large dissemination tree that reaches many new nodes in a single time step, short adoption paths, and small structural viralities. The maxima in the longest-path length and structural virality indicate that a spreading process persists longer (and hence travels farther from a source node) when nodes have fewer neighbors on average (but still have {a mean degree that is large enough} for an infodemic to occur).

Varying the receptiveness parameter $c$ (see \Cref{fig:poisson-varyc}) yields similar dissemination-tree statistics as varying the expected mean degree. We again observe maxima in the longest-path length and the structural virality{. These maxima likely arise because larger receptiveness values favoring faster content spread, although the longest-path length and structural virality both have large standard deviations in this experiment.} We observe evidence of a phase transition in the total number of content shares at the value of {the receptiveness parameter} $c$ that we predicted {in} our analysis in \Cref{sec:infodemic-analysis}. There also appears to be an accompanying phase transition in the dissemination-tree width. {In these simulations, the content state is $x_0 = 0.5$. We observe qualitatively similar behavior with content states (e.g., $x_0 = 0.15$) that are closer to the boundary of opinion space but still in the infodemic regime. In our repository (see \url{https://github.com/hzinnbrooks/bounded-confidence-spreading-process/supplementary-figures}), we include analogous figures to \Cref{fig:poisson-varyk} and \Cref{fig:poisson-varyc} for such content states.}

We also examine the impact of the content state on our dissemination-tree statistics for two different initial opinion distributions (see \Cref{eqn:probshare}). 
 In \Cref{fig:poisson-varyx0}, we show results of simulations in we draw the initial node opinions from a uniform distribution on the interval $(0,1)$. 
In \Cref{fig:poisson-varyx0-Gaussian}, we show results of simulations in which we draw the initial node opinions {from} $\mathbb{R}$ {using} a Gaussian distribution with mean $0.5$ and standard deviation $0.1$.
 As expected, we observe smaller values of all four summary statistics when the content state lies in the tails of a opinion distribution $\phi(x)$ (or near the boundary of the distribution interval for the uniform distribution). 
 We observe maxima in the total number of content shares and dissemination-tree width at the distribution mean (and also for a very large interval around it for the uniform distribution).
 For the Gaussian distribution, the content states that yield the largest longest-path lengths and structural viralities occur away from the mean (and are symmetric), indicating that content can spread farther from a
 source node in these situations then when the content state equals the mean initial opinion. These dissemination trees are longer and narrow than when the content state equals the mean initial opinion. If the content state is too extreme, it spreads very little. 


\section{Conclusions and discussion} \label{sec:discussion}

We examined ``infodemics'', in the form of large content-spreading cascades, in a bounded-confidence content-spreading model on configuration-model networks. 
To do this, we defined an ``opinion reproduction number" that is analogous to the basic reproduction number of disease dynamics. 
{By examining whether the opinion of the content and the receptiveness of individuals yield an opinion reproduction number that is larger than $1$, we investigated when content propagates widely in a network of agents and when it does not.} 

We quantified the size and structure of content spread by measuring several properties of dissemination trees --- tree size, tree width, the length of the longest adoption paths, and structural virality --- and we thereby illustrated how network structure and spreading-model parameters affect how content spreads. We found that larger networks, larger expected mean degrees, larger receptiveness, and content states near the mean of opinion distributions all promote larger total numbers of content shares, including the possibility of sharing the content to many new nodes in a single time step. Additionally, when the expected mean degree and receptiveness levels are small but place the spreading process above the infodemic threshold, there can be longer dissemination trees than when the expected mean degree or receptiveness are larger, even though the total number of content shares is smaller.
This indicates that content is spreading farther but less widely from a source agent (i.e., an agent that originally posts a piece of content).

There are many interesting ways to extend our work. A particularly relevant extension is to examine the effects of purposeful choices of source nodes, such as to try to promote influence maximization {\cite{peng2018,kkt2003}.}
One can also examine the effects of competing social contagions \cite{gleeson2014}, which perhaps are spread by agents with different political perspectives or agendas.
Another worthwhile future direction is to adapt the recently developed ``distributed reproduction numbers" \cite{she2023} from disease spread to content spread.
It is also desirable to extend our study of content-spreading dynamics to other network structures. One key direction is to consider more complicated network models (such as generalizations of configuration models that incorporate various types of heterogeneities \cite{melnik2014}) and real-world networks. It is also natural to extend our investigation to more complicated types of networks, such as multilayer networks (which allow multiple types of social connections and communication channels) \cite{kivela2014}, hypergraphs (which allow simultaneous interactions between three or more agents) \cite{battiston2020}, and adaptive networks (e.g., to incorporate changing relationship structures such as ``unfollowing'' or ``unfriending'' on social-media platforms) \cite{berner2023}.

{Other viable extensions of our model directly exploit the link that it establishes between opinion dynamics and percolation processes. It seems particularly exciting to study scenarios with rich interplays between
the opinion of content and the opinions of agents. For example, one can incorporate mutations of content opinions \cite{friggeri2014,adamic2016} through averaging the the opinion of content with the opinions of the agents that spread it.
   A related extension entails allowing agent opinions to evolve with time through their interactions with each other (as in classical BCMs, such as the Deffuant--Weisbuch model \cite{deffuant2000mixing}) while content with a fixed opinion 
  spreads on a network. It will be fascinating to explore how these extensions affect the properties of infodemic thresholds and spreading patterns.}


\section*{Software}

We performed computations and constructed visualizations using MATLAB. 
The code to reproduce our numerical simulations is available at \url{https://github.com/hzinnbrooks/bounded-confidence-spreading-process}. 


\begin{acknowledgments}

    We thank {Abhinav Chand, Jonas Juul,} and two anonymous referees for helpful comments. HZB was funded in part by the National Science Foundation (grant number DMS-2109239) through their program on Applied Mathematics. MAP acknowledges ﬁnancial support from the National Science Foundation (grant number 1922952) through their program on Algorithms for Threat Detection (ATD). MAP thanks David Campbell for his friendship and many discussions over the years.


\end{acknowledgments}


\appendix

\section{A few words about David Campbell (by Mason A. Porter)} \label{david}

It is a great honor to contribute to \emph{Chaos}'s special issue in celebration of David Campbell's 80th birthday. David is one of my favorite people in science, and I deeply appreciate being asked to contribute an article to this.Festschrift. David Campbell's research contributions and, especially, his foundational role in \emph{Chaos} are well-known. I could write about that, but I will leave that to others. What is especially important is that David is a great human being. He has always been fair to me and he has been very good to me, as he has been to so many others. 

I first interacted with David when I was a postdoc at Georgia Tech. I was naively trying to publish a modified version of my doctoral thesis \cite{porter2002} as a review article, and I submitted it to the journal \emph{Physics Reports}. David, who I {had not yet met} in person, was the handling editor of my submission. My paper was rejected, but what stood out in that process was the carefulness and decency in which David handled the process. {All that} I want with a manuscript submission --- and all {that} anyone should ever want with a manuscript submission, as both a necessary and sufficient condition --- is to be treated fairly and with dignity. David saw a teaching opportunity, and he could see that I had things to {learn; he} went beyond the call of duty and even asked the referee to disclose his identity to help with that teaching. By contrast, as we all see repeatedly, many journal editors seem to be bean counters, rather than actual editors. {The rejection of my paper} wasn't my ideal outcome, but it was the correct outcome, and I learned a lot from it. In my own editing roles, I have tried to draw on the approach and lessons that David exemplified in this story.

David continued looking out for me, as he has looked out for many others. Having --- I think? --- still not yet met in person, David invited me to give a talk as part of a special session at the 2005 American Physical Society (APS) March Meeting about the celebrated Fermi--Pasta--Ulam--Tsingou (FPUT) problem and {to} contribute an article about the dynamics of Bose--Einstein condensates~\cite{porter2005} to a special issue in \emph{Chaos} about the FPUT problem. David's talk had a pop-music reference (which was about the Eagles, if memory serves), and he mentioned that he figured that I'd appreciate the music reference. (He was right.) Four of us in the special session eventually coauthored a popular article about the FPUT problem \cite{porter2009}. I eventually wrote a variety of papers on the FPUT problem (and on {related} problems), including one \cite{nelson2018} that turned out to be refereed by David and one of his students. It was a particularly fair and thorough referee report --- of the type that scientists dream of getting {and} that indicates that sometimes the reviewing process actually works the way that it it supposed to --- that my coauthors were spending time trying to address. I remember {e-mailing David} during our revision process to tell him about our paper{, which concerns heterogeneities in FPUT lattices, because I figured that he would be interested it.} My e-mail to David either was about something else entirely or was because I had just seen a paper of his on a related topic. In any event, without {having any notion that David was our} referee, I told him in my e-mail about our paper and that we were currently dealing with a tough-but-fair referee report. I found out from David's response that he and his student had written that report. Once again, David was doing things in a way that helps authors while upholding rigorous standards. In other words, David once again did things in the way that they are supposed to be done.

A recurring theme in my stories is David looking out for a junior scientist, {treating people fairly while simultaneously ensuring rigorous standards,} and doing things the right way in scientific editing and reviewing. We all have reviewing horror stories{,} and publishing can cause many frustrations. {However,} some scientists consistently do things the right way, and David is one of them.

The best complement that somebody like me who comes from Jewish heritage can pay to another person is to call them a ``mensch" (which, essentially, is {saying that somebody is} a decent human being, as they are a person of integrity and honor). David Campbell is a mensch.

Happy birthday, David!



\begin{thebibliography}{100}%
\makeatletter
\providecommand \@ifxundefined [1]{%
 \@ifx{#1\undefined}
}%
\providecommand \@ifnum [1]{%
 \ifnum #1\expandafter \@firstoftwo
 \else \expandafter \@secondoftwo
 \fi
}%
\providecommand \@ifx [1]{%
 \ifx #1\expandafter \@firstoftwo
 \else \expandafter \@secondoftwo
 \fi
}%
\providecommand \natexlab [1]{#1}%
\providecommand \enquote  [1]{``#1''}%
\providecommand \bibnamefont  [1]{#1}%
\providecommand \bibfnamefont [1]{#1}%
\providecommand \citenamefont [1]{#1}%
\providecommand \href@noop [0]{\@secondoftwo}%
\providecommand \href [0]{\begingroup \@sanitize@url \@href}%
\providecommand \@href[1]{\@@startlink{#1}\@@href}%
\providecommand \@@href[1]{\endgroup#1\@@endlink}%
\providecommand \@sanitize@url [0]{\catcode `\\12\catcode `\$12\catcode
  `\&12\catcode `\#12\catcode `\^12\catcode `\_12\catcode `\%12\relax}%
\providecommand \@@startlink[1]{}%
\providecommand \@@endlink[0]{}%
\providecommand \url  [0]{\begingroup\@sanitize@url \@url }%
\providecommand \@url [1]{\endgroup\@href {#1}{\urlprefix }}%
\providecommand \urlprefix  [0]{URL }%
\providecommand \Eprint [0]{\href }%
\providecommand \doibase [0]{http://dx.doi.org/}%
\providecommand \selectlanguage [0]{\@gobble}%
\providecommand \bibinfo  [0]{\@secondoftwo}%
\providecommand \bibfield  [0]{\@secondoftwo}%
\providecommand \translation [1]{[#1]}%
\providecommand \BibitemOpen [0]{}%
\providecommand \bibitemStop [0]{}%
\providecommand \bibitemNoStop [0]{.\EOS\space}%
\providecommand \EOS [0]{\spacefactor3000\relax}%
\providecommand \BibitemShut  [1]{\csname bibitem#1\endcsname}%
\let\auto@bib@innerbib\@empty
\bibitem [{\citenamefont {Goel}\ \emph {et~al.}(2016)\citenamefont {Goel},
  \citenamefont {Anderson}, \citenamefont {Hofman},\ and\ \citenamefont
  {Watts}}]{goel2016structural}%
  \BibitemOpen
  \bibfield  {author} {\bibinfo {author} {\bibfnamefont {S.}~\bibnamefont
  {Goel}}, \bibinfo {author} {\bibfnamefont {A.}~\bibnamefont {Anderson}},
  \bibinfo {author} {\bibfnamefont {J.}~\bibnamefont {Hofman}}, \ and\ \bibinfo
  {author} {\bibfnamefont {D.~J.}\ \bibnamefont {Watts}},\ }\bibfield  {title}
  {\enquote {\bibinfo {title} {The structural virality of online diffusion},}\
  }\href@noop {} {\bibfield  {journal} {\bibinfo  {journal} {Management
  Science}\ }\textbf {\bibinfo {volume} {62}},\ \bibinfo {pages} {180--196}
  (\bibinfo {year} {2016})}\BibitemShut {NoStop}%
\bibitem [{\citenamefont {Juul}\ and\ \citenamefont
  {Ugander}(2021)}]{juul2022}%
  \BibitemOpen
  \bibfield  {author} {\bibinfo {author} {\bibfnamefont {J.~L.}\ \bibnamefont
  {Juul}}\ and\ \bibinfo {author} {\bibfnamefont {J.}~\bibnamefont {Ugander}},\
  }\bibfield  {title} {\enquote {\bibinfo {title} {Comparing information
  diffusion mechanisms by matching on cascade size},}\ }\href@noop {}
  {\bibfield  {journal} {\bibinfo  {journal} {Proceedings of the National
  Academy of Sciences of the United States of America}\ }\textbf {\bibinfo
  {volume} {118}},\ \bibinfo {eid} {e2100786118} (\bibinfo {year}
  {2021})}\BibitemShut {NoStop}%
\bibitem [{\citenamefont {Eysenbach}(2002)}]{eysenbach2002}%
  \BibitemOpen
  \bibfield  {author} {\bibinfo {author} {\bibfnamefont {G.}~\bibnamefont
  {Eysenbach}},\ }\bibfield  {title} {\enquote {\bibinfo {title}
  {Infodemiology: {T}he epidemiology of (mis)information},}\ }\href@noop {}
  {\bibfield  {journal} {\bibinfo  {journal} {The American Journal of
  Medicine}\ }\textbf {\bibinfo {volume} {13}},\ \bibinfo {pages} {763--765}
  (\bibinfo {year} {2002})}\BibitemShut {NoStop}%
\bibitem [{\citenamefont {Briand}\ \emph {et~al.}(2021)\citenamefont {Briand},
  \citenamefont {Cinelli}, \citenamefont {Nguyen}, \citenamefont {Lewis},
  \citenamefont {Prybylski}, \citenamefont {Valensise}, \citenamefont
  {Colizza}, \citenamefont {Tozzi}, \citenamefont {Perra}, \citenamefont
  {Baronchelli}, \citenamefont {Tizzoni}, \citenamefont {Zollo}, \citenamefont
  {Scala}, \citenamefont {Purnat}, \citenamefont {Czerniak}, \citenamefont
  {Kucharski}, \citenamefont {Tshangela}, \citenamefont {Zhou},\ and\
  \citenamefont {Quattrociocchi}}]{briand2021}%
  \BibitemOpen
  \bibfield  {author} {\bibinfo {author} {\bibfnamefont {S.~C.}\ \bibnamefont
  {Briand}}, \bibinfo {author} {\bibfnamefont {M.}~\bibnamefont {Cinelli}},
  \bibinfo {author} {\bibfnamefont {T.}~\bibnamefont {Nguyen}}, \bibinfo
  {author} {\bibfnamefont {R.}~\bibnamefont {Lewis}}, \bibinfo {author}
  {\bibfnamefont {D.}~\bibnamefont {Prybylski}}, \bibinfo {author}
  {\bibfnamefont {C.~M.}\ \bibnamefont {Valensise}}, \bibinfo {author}
  {\bibfnamefont {V.}~\bibnamefont {Colizza}}, \bibinfo {author} {\bibfnamefont
  {A.~E.}\ \bibnamefont {Tozzi}}, \bibinfo {author} {\bibfnamefont
  {N.}~\bibnamefont {Perra}}, \bibinfo {author} {\bibfnamefont
  {A.}~\bibnamefont {Baronchelli}}, \bibinfo {author} {\bibfnamefont
  {M.}~\bibnamefont {Tizzoni}}, \bibinfo {author} {\bibfnamefont
  {F.}~\bibnamefont {Zollo}}, \bibinfo {author} {\bibfnamefont
  {A.}~\bibnamefont {Scala}}, \bibinfo {author} {\bibfnamefont
  {T.}~\bibnamefont {Purnat}}, \bibinfo {author} {\bibfnamefont
  {C.}~\bibnamefont {Czerniak}}, \bibinfo {author} {\bibfnamefont {A.~J.}\
  \bibnamefont {Kucharski}}, \bibinfo {author} {\bibfnamefont {A.}~\bibnamefont
  {Tshangela}}, \bibinfo {author} {\bibfnamefont {L.}~\bibnamefont {Zhou}}, \
  and\ \bibinfo {author} {\bibfnamefont {W.}~\bibnamefont {Quattrociocchi}},\
  }\bibfield  {title} {\enquote {\bibinfo {title} {Infodemics: {A} new
  challenge for public health},}\ }\href@noop {} {\bibfield  {journal}
  {\bibinfo  {journal} {Cell}\ }\textbf {\bibinfo {volume} {184}},\ \bibinfo
  {pages} {6010--6014} (\bibinfo {year} {2021})}\BibitemShut {NoStop}%
\bibitem [{\citenamefont {Zielinski}(2021)}]{zielinski2021}%
  \BibitemOpen
  \bibfield  {author} {\bibinfo {author} {\bibfnamefont {C.}~\bibnamefont
  {Zielinski}},\ }\bibfield  {title} {\enquote {\bibinfo {title} {Infodemics
  and infodemiology: {A} short history, a long future},}\ }\href@noop {}
  {\bibfield  {journal} {\bibinfo  {journal} {Pan American Journal of Public
  Health}\ }\textbf {\bibinfo {volume} {45}},\ \bibinfo {eid} {e40} (\bibinfo
  {year} {2021})}\BibitemShut {NoStop}%
\bibitem [{\citenamefont {Kiss}, \citenamefont {Miller},\ and\ \citenamefont
  {Simon}(2017)}]{kiss2017mathematics}%
  \BibitemOpen
  \bibfield  {author} {\bibinfo {author} {\bibfnamefont {I.~Z.}\ \bibnamefont
  {Kiss}}, \bibinfo {author} {\bibfnamefont {J.~C.}\ \bibnamefont {Miller}}, \
  and\ \bibinfo {author} {\bibfnamefont {P.~L.}\ \bibnamefont {Simon}},\
  }\href@noop {} {\emph {\bibinfo {title} {Mathematics of {E}pidemics on
  {N}etworks: From {E}xact to {A}pproximate {M}odels}}}\ (\bibinfo  {publisher}
  {Springer},\ \bibinfo {address} {Cham, Switzerland},\ \bibinfo {year}
  {2017})\BibitemShut {NoStop}%
\bibitem [{\citenamefont {Brauer}, \citenamefont {Castillo-Chavez},\ and\
  \citenamefont {Feng}(2019)}]{brauer2019}%
  \BibitemOpen
  \bibfield  {author} {\bibinfo {author} {\bibfnamefont {F.}~\bibnamefont
  {Brauer}}, \bibinfo {author} {\bibfnamefont {C.}~\bibnamefont
  {Castillo-Chavez}}, \ and\ \bibinfo {author} {\bibfnamefont {Z.}~\bibnamefont
  {Feng}},\ }\href@noop {} {\emph {\bibinfo {title} {Mathematical Models in
  Epidemiology}}}\ (\bibinfo  {publisher} {Springer},\ \bibinfo {address}
  {Heidelberg, Germany},\ \bibinfo {year} {2019})\BibitemShut {NoStop}%
\bibitem [{\citenamefont {Gleeson}\ \emph {et~al.}(2021)\citenamefont
  {Gleeson}, \citenamefont {Onaga}, \citenamefont {Fennell}, \citenamefont
  {Cotter}, \citenamefont {Burke},\ and\ \citenamefont
  {O'Sullivan}}]{gleeson2021}%
  \BibitemOpen
  \bibfield  {author} {\bibinfo {author} {\bibfnamefont {J.~P.}\ \bibnamefont
  {Gleeson}}, \bibinfo {author} {\bibfnamefont {T.}~\bibnamefont {Onaga}},
  \bibinfo {author} {\bibfnamefont {P.}~\bibnamefont {Fennell}}, \bibinfo
  {author} {\bibfnamefont {J.}~\bibnamefont {Cotter}}, \bibinfo {author}
  {\bibfnamefont {R.}~\bibnamefont {Burke}}, \ and\ \bibinfo {author}
  {\bibfnamefont {D.~J.~P.}\ \bibnamefont {O'Sullivan}},\ }\bibfield  {title}
  {\enquote {\bibinfo {title} {Branching process descriptions of information
  cascades on {T}witter},}\ }\href@noop {} {\bibfield  {journal} {\bibinfo
  {journal} {Journal of Complex Networks}\ }\textbf {\bibinfo {volume} {8}},\
  \bibinfo {eid} {cnab002} (\bibinfo {year} {2021})}\BibitemShut {NoStop}%
\bibitem [{\citenamefont {Bak-Coleman}\ \emph {et~al.}(2021)\citenamefont
  {Bak-Coleman}, \citenamefont {Alfano}, \citenamefont {Barfuss}, \citenamefont
  {Bergstrom}, \citenamefont {Centeno}, \citenamefont {Couzin}, \citenamefont
  {Donges}, \citenamefont {Galesic}, \citenamefont {Gersick}, \citenamefont
  {Jacquet}, \citenamefont {Kao}, \citenamefont {Moran}, \citenamefont
  {Romanczuk}, \citenamefont {Rubenstein}, \citenamefont {Tombak},
  \citenamefont {Van~Bavel},\ and\ \citenamefont {Weber}}]{bak2021}%
  \BibitemOpen
  \bibfield  {author} {\bibinfo {author} {\bibfnamefont {J.~B.}\ \bibnamefont
  {Bak-Coleman}}, \bibinfo {author} {\bibfnamefont {M.}~\bibnamefont {Alfano}},
  \bibinfo {author} {\bibfnamefont {W.}~\bibnamefont {Barfuss}}, \bibinfo
  {author} {\bibfnamefont {C.~T.}\ \bibnamefont {Bergstrom}}, \bibinfo {author}
  {\bibfnamefont {M.~A.}\ \bibnamefont {Centeno}}, \bibinfo {author}
  {\bibfnamefont {I.~D.}\ \bibnamefont {Couzin}}, \bibinfo {author}
  {\bibfnamefont {J.~F.}\ \bibnamefont {Donges}}, \bibinfo {author}
  {\bibfnamefont {M.}~\bibnamefont {Galesic}}, \bibinfo {author} {\bibfnamefont
  {A.~S.}\ \bibnamefont {Gersick}}, \bibinfo {author} {\bibfnamefont
  {J.}~\bibnamefont {Jacquet}}, \bibinfo {author} {\bibfnamefont {A.~B.}\
  \bibnamefont {Kao}}, \bibinfo {author} {\bibfnamefont {R.~E.}\ \bibnamefont
  {Moran}}, \bibinfo {author} {\bibfnamefont {P.}~\bibnamefont {Romanczuk}},
  \bibinfo {author} {\bibfnamefont {D.~I.}\ \bibnamefont {Rubenstein}},
  \bibinfo {author} {\bibfnamefont {K.~J.}\ \bibnamefont {Tombak}}, \bibinfo
  {author} {\bibfnamefont {J.~J.}\ \bibnamefont {Van~Bavel}}, \ and\ \bibinfo
  {author} {\bibfnamefont {E.~U.}\ \bibnamefont {Weber}},\ }\bibfield  {title}
  {\enquote {\bibinfo {title} {Stewardship of global collective behavior},}\
  }\href@noop {} {\bibfield  {journal} {\bibinfo  {journal} {Proc. Nat. Acad.
  Sci. USA}\ }\textbf {\bibinfo {volume} {118}},\ \bibinfo {eid} {e2025764118}
  (\bibinfo {year} {2021})}\BibitemShut {NoStop}%
\bibitem [{\citenamefont {Guille}\ \emph {et~al.}(2013)\citenamefont {Guille},
  \citenamefont {Hacid}, \citenamefont {Favre},\ and\ \citenamefont
  {Zighed}}]{guille2013information}%
  \BibitemOpen
  \bibfield  {author} {\bibinfo {author} {\bibfnamefont {A.}~\bibnamefont
  {Guille}}, \bibinfo {author} {\bibfnamefont {H.}~\bibnamefont {Hacid}},
  \bibinfo {author} {\bibfnamefont {C.}~\bibnamefont {Favre}}, \ and\ \bibinfo
  {author} {\bibfnamefont {D.~A.}\ \bibnamefont {Zighed}},\ }\bibfield  {title}
  {\enquote {\bibinfo {title} {Information diffusion in online social networks:
  {A} survey},}\ }\href@noop {} {\bibfield  {journal} {\bibinfo  {journal} {ACM
  Sigmod Record}\ }\textbf {\bibinfo {volume} {42}},\ \bibinfo {pages} {17--28}
  (\bibinfo {year} {2013})}\BibitemShut {NoStop}%
\bibitem [{\citenamefont {K\"{u}mpel}, \citenamefont {Karnowski},\ and\
  \citenamefont {Keyling}(2015)}]{kumpel2015}%
  \BibitemOpen
  \bibfield  {author} {\bibinfo {author} {\bibfnamefont {A.~S.}\ \bibnamefont
  {K\"{u}mpel}}, \bibinfo {author} {\bibfnamefont {V.}~\bibnamefont
  {Karnowski}}, \ and\ \bibinfo {author} {\bibfnamefont {T.}~\bibnamefont
  {Keyling}},\ }\bibfield  {title} {\enquote {\bibinfo {title} {News sharing in
  social media: {A} review of current research on news sharing users, content,
  and networks},}\ }\href@noop {} {\bibfield  {journal} {\bibinfo  {journal}
  {Social Media + Society}\ }\textbf {\bibinfo {volume} {1}},\ \bibinfo {eid}
  {2056305115610141} (\bibinfo {year} {2015})}\BibitemShut {NoStop}%
\bibitem [{\citenamefont {{Muhammed T}}\ and\ \citenamefont
  {Mathew}(2022)}]{muhammed2022}%
  \BibitemOpen
  \bibfield  {author} {\bibinfo {author} {\bibfnamefont {S.}~\bibnamefont
  {{Muhammed T}}}\ and\ \bibinfo {author} {\bibfnamefont {S.~K.}\ \bibnamefont
  {Mathew}},\ }\bibfield  {title} {\enquote {\bibinfo {title} {The disaster of
  misinformation: {A} review of research in social media},}\ }\href@noop {}
  {\bibfield  {journal} {\bibinfo  {journal} {International Journal of Data
  Science and Analytics}\ }\textbf {\bibinfo {volume} {13}},\ \bibinfo {pages}
  {271--285} (\bibinfo {year} {2022})}\BibitemShut {NoStop}%
\bibitem [{\citenamefont {Friggeri}\ \emph {et~al.}(2014)\citenamefont
  {Friggeri}, \citenamefont {Adamic}, \citenamefont {Eckles},\ and\
  \citenamefont {Cheng}}]{friggeri2014}%
  \BibitemOpen
  \bibfield  {author} {\bibinfo {author} {\bibfnamefont {A.}~\bibnamefont
  {Friggeri}}, \bibinfo {author} {\bibfnamefont {L.}~\bibnamefont {Adamic}},
  \bibinfo {author} {\bibfnamefont {D.}~\bibnamefont {Eckles}}, \ and\ \bibinfo
  {author} {\bibfnamefont {J.}~\bibnamefont {Cheng}},\ }\bibfield  {title}
  {\enquote {\bibinfo {title} {Rumor cascades},}\ }\href@noop {} {\bibfield
  {journal} {\bibinfo  {journal} {Proceedings of the International AAAI
  Conference on Web and Social Media}\ }\textbf {\bibinfo {volume} {8}},\
  \bibinfo {pages} {101--110} (\bibinfo {year} {2014})}\BibitemShut {NoStop}%
\bibitem [{\citenamefont {Starbird}(2019)}]{starbird2019}%
  \BibitemOpen
  \bibfield  {author} {\bibinfo {author} {\bibfnamefont {K.}~\bibnamefont
  {Starbird}},\ }\bibfield  {title} {\enquote {\bibinfo {title}
  {Disinformation's spread: {B}ots, trolls and all of us},}\ }\href@noop {}
  {\bibfield  {journal} {\bibinfo  {journal} {Nature}\ }\textbf {\bibinfo
  {volume} {571}},\ \bibinfo {pages} {449--450} (\bibinfo {year}
  {2019})}\BibitemShut {NoStop}%
\bibitem [{\citenamefont {Yang}\ \emph {et~al.}(2023)\citenamefont {Yang},
  \citenamefont {Varol}, \citenamefont {Nwala}, \citenamefont
  {Sayyadiharikandeh}, \citenamefont {Ferrara}, \citenamefont {Flammini},\ and\
  \citenamefont {Menczer}}]{yang2023}%
  \BibitemOpen
  \bibfield  {author} {\bibinfo {author} {\bibfnamefont {K.-C.}\ \bibnamefont
  {Yang}}, \bibinfo {author} {\bibfnamefont {O.}~\bibnamefont {Varol}},
  \bibinfo {author} {\bibfnamefont {A.~C.}\ \bibnamefont {Nwala}}, \bibinfo
  {author} {\bibfnamefont {M.}~\bibnamefont {Sayyadiharikandeh}}, \bibinfo
  {author} {\bibfnamefont {E.}~\bibnamefont {Ferrara}}, \bibinfo {author}
  {\bibfnamefont {A.}~\bibnamefont {Flammini}}, \ and\ \bibinfo {author}
  {\bibfnamefont {F.}~\bibnamefont {Menczer}},\ }\bibfield  {title} {\enquote
  {\bibinfo {title} {Social bots: {D}etection and challenges},}\ }\href@noop {}
  {\bibfield  {journal} {\bibinfo  {journal} {arXiv preprint arXiv:2312.17423}\
  } (\bibinfo {year} {2023})},\ \bibinfo {note} {[to appear as a chapter in
  \emph{Handbook of Computational Social Science} (edited by Taha Yasseri),
  forthcoming in 2024, Edward Elgar Publishing Ltd., Cheltenham,
  UK}\BibitemShut {NoStop}%
\bibitem [{\citenamefont {L\'{o}pez}, \citenamefont {Pastor-Galindo},\ and\
  \citenamefont {Ruip\'{e}rez-Valiente}(2024)}]{buitrago2024}%
  \BibitemOpen
  \bibfield  {author} {\bibinfo {author} {\bibfnamefont {A.~B.}\ \bibnamefont
  {L\'{o}pez}}, \bibinfo {author} {\bibfnamefont {J.}~\bibnamefont
  {Pastor-Galindo}}, \ and\ \bibinfo {author} {\bibfnamefont {J.~A.}\
  \bibnamefont {Ruip\'{e}rez-Valiente}},\ }\bibfield  {title} {\enquote
  {\bibinfo {title} {Frameworks, modeling and simulations of misinformation and
  disinformation: {A} systematic literature review},}\ }\href@noop {}
  {\bibfield  {journal} {\bibinfo  {journal} {arXiv preprint arXiv:2406.09343}\
  } (\bibinfo {year} {2024})}\BibitemShut {NoStop}%
\bibitem [{\citenamefont {Hartmann}\ \emph {et~al.}(2024)\citenamefont
  {Hartmann}, \citenamefont {Pohlmann}, \citenamefont {Wang},\ and\
  \citenamefont {Berendt}}]{echo2024}%
  \BibitemOpen
  \bibfield  {author} {\bibinfo {author} {\bibfnamefont {D.}~\bibnamefont
  {Hartmann}}, \bibinfo {author} {\bibfnamefont {L.}~\bibnamefont {Pohlmann}},
  \bibinfo {author} {\bibfnamefont {S.~M.}\ \bibnamefont {Wang}}, \ and\
  \bibinfo {author} {\bibfnamefont {B.}~\bibnamefont {Berendt}},\ }\bibfield
  {title} {\enquote {\bibinfo {title} {A systematic review of echo chamber
  research: {C}omparative analysis of conceptualizations, operationalizations,
  and varying outcomes},}\ }\href@noop {} {\bibfield  {journal} {\bibinfo
  {journal} {arXiv preprint arXiv:2407.06631}\ } (\bibinfo {year}
  {2024})}\BibitemShut {NoStop}%
\bibitem [{\citenamefont {Pennycook}\ \emph {et~al.}(2021)\citenamefont
  {Pennycook}, \citenamefont {Epstein}, \citenamefont {Mosleh}, \citenamefont
  {Arechar}, \citenamefont {Eckles},\ and\ \citenamefont
  {Rand}}]{pennycook2021}%
  \BibitemOpen
  \bibfield  {author} {\bibinfo {author} {\bibfnamefont {G.}~\bibnamefont
  {Pennycook}}, \bibinfo {author} {\bibfnamefont {Z.}~\bibnamefont {Epstein}},
  \bibinfo {author} {\bibfnamefont {M.}~\bibnamefont {Mosleh}}, \bibinfo
  {author} {\bibfnamefont {A.~A.}\ \bibnamefont {Arechar}}, \bibinfo {author}
  {\bibfnamefont {D.}~\bibnamefont {Eckles}}, \ and\ \bibinfo {author}
  {\bibfnamefont {D.~G.}\ \bibnamefont {Rand}},\ }\bibfield  {title} {\enquote
  {\bibinfo {title} {Shifting attention to accuracy can reduce misinformation
  online},}\ }\href@noop {} {\bibfield  {journal} {\bibinfo  {journal}
  {Nature}\ }\textbf {\bibinfo {volume} {592}},\ \bibinfo {pages} {590--595}
  (\bibinfo {year} {2021})}\BibitemShut {NoStop}%
\bibitem [{\citenamefont {Baek}\ \emph {et~al.}(2023)\citenamefont {Baek},
  \citenamefont {Hyon}, \citenamefont {L\'{o}pez}, \citenamefont {Porter},\
  and\ \citenamefont {Parkinson}}]{baek2023}%
  \BibitemOpen
  \bibfield  {author} {\bibinfo {author} {\bibfnamefont {E.~C.}\ \bibnamefont
  {Baek}}, \bibinfo {author} {\bibfnamefont {R.}~\bibnamefont {Hyon}}, \bibinfo
  {author} {\bibfnamefont {K.}~\bibnamefont {L\'{o}pez}}, \bibinfo {author}
  {\bibfnamefont {M.~A.}\ \bibnamefont {Porter}}, \ and\ \bibinfo {author}
  {\bibfnamefont {C.}~\bibnamefont {Parkinson}},\ }\bibfield  {title} {\enquote
  {\bibinfo {title} {Perceived community alignment increases information
  sharing},}\ }\href@noop {} {\bibfield  {journal} {\bibinfo  {journal} {arXiv
  preprint arXiv:2304.13796}\ } (\bibinfo {year} {2023})}\BibitemShut {NoStop}%
\bibitem [{\citenamefont {Aral}, \citenamefont {Muchnik},\ and\ \citenamefont
  {Sundararajan}(2009)}]{aral2009}%
  \BibitemOpen
  \bibfield  {author} {\bibinfo {author} {\bibfnamefont {S.}~\bibnamefont
  {Aral}}, \bibinfo {author} {\bibfnamefont {L.}~\bibnamefont {Muchnik}}, \
  and\ \bibinfo {author} {\bibfnamefont {A.}~\bibnamefont {Sundararajan}},\
  }\bibfield  {title} {\enquote {\bibinfo {title} {Distinguishing
  influence-based contagion from homophily-driven diffusion in dynamic
  networks},}\ }\href@noop {} {\bibfield  {journal} {\bibinfo  {journal}
  {Proceedings of the National Academy of Sciences of the United States of
  America}\ }\textbf {\bibinfo {volume} {106}},\ \bibinfo {pages}
  {21544--21549} (\bibinfo {year} {2009})}\BibitemShut {NoStop}%
\bibitem [{\citenamefont {Christakis}\ and\ \citenamefont
  {Fowler}(2013)}]{christakis2013}%
  \BibitemOpen
  \bibfield  {author} {\bibinfo {author} {\bibfnamefont {N.~A.}\ \bibnamefont
  {Christakis}}\ and\ \bibinfo {author} {\bibfnamefont {J.~H.}\ \bibnamefont
  {Fowler}},\ }\bibfield  {title} {\enquote {\bibinfo {title} {Social contagion
  theory: {E}xamining dynamic social networks and human behavior},}\
  }\href@noop {} {\bibfield  {journal} {\bibinfo  {journal} {Statistics In
  Medicine}\ }\textbf {\bibinfo {volume} {32}},\ \bibinfo {pages} {556--577}
  (\bibinfo {year} {2013})}\BibitemShut {NoStop}%
\bibitem [{\citenamefont {Borge-Holthoefer}\ \emph {et~al.}(2013)\citenamefont
  {Borge-Holthoefer}, \citenamefont {Ba{\~n}os}, \citenamefont
  {Gonz\'{a}lez-Bail\'{o}n},\ and\ \citenamefont {Moreno}}]{borge2013}%
  \BibitemOpen
  \bibfield  {author} {\bibinfo {author} {\bibfnamefont {J.}~\bibnamefont
  {Borge-Holthoefer}}, \bibinfo {author} {\bibfnamefont {R.~A.}\ \bibnamefont
  {Ba{\~n}os}}, \bibinfo {author} {\bibfnamefont {S.}~\bibnamefont
  {Gonz\'{a}lez-Bail\'{o}n}}, \ and\ \bibinfo {author} {\bibfnamefont
  {Y.}~\bibnamefont {Moreno}},\ }\bibfield  {title} {\enquote {\bibinfo {title}
  {Cascading behaviour in complex socio-technical networks},}\ }\href@noop {}
  {\bibfield  {journal} {\bibinfo  {journal} {Journal of Complex Networks}\
  }\textbf {\bibinfo {volume} {1}},\ \bibinfo {pages} {3--24} (\bibinfo {year}
  {2013})}\BibitemShut {NoStop}%
\bibitem [{\citenamefont {Yu}\ \emph {et~al.}(2020)\citenamefont {Yu},
  \citenamefont {Huang}, \citenamefont {Liu},\ and\ \citenamefont
  {Tan}}]{yu2020}%
  \BibitemOpen
  \bibfield  {author} {\bibinfo {author} {\bibfnamefont {Y.}~\bibnamefont
  {Yu}}, \bibinfo {author} {\bibfnamefont {S.}~\bibnamefont {Huang}}, \bibinfo
  {author} {\bibfnamefont {Y.}~\bibnamefont {Liu}}, \ and\ \bibinfo {author}
  {\bibfnamefont {Y.}~\bibnamefont {Tan}},\ }\bibfield  {title} {\enquote
  {\bibinfo {title} {Emotions in online content diffusion},}\ }\href@noop {}
  {\bibfield  {journal} {\bibinfo  {journal} {arXiv preprint arXiv:2011.09003}\
  } (\bibinfo {year} {2020})}\BibitemShut {NoStop}%
\bibitem [{\citenamefont {Pr\"{o}llochs}, \citenamefont {B\"{a}r},\ and\
  \citenamefont {Feuerriegel}(2021)}]{prollochs2021}%
  \BibitemOpen
  \bibfield  {author} {\bibinfo {author} {\bibfnamefont {N.}~\bibnamefont
  {Pr\"{o}llochs}}, \bibinfo {author} {\bibfnamefont {D.}~\bibnamefont
  {B\"{a}r}}, \ and\ \bibinfo {author} {\bibfnamefont {S.}~\bibnamefont
  {Feuerriegel}},\ }\bibfield  {title} {\enquote {\bibinfo {title} {Emotions in
  online rumor diffusion},}\ }\href@noop {} {\bibfield  {journal} {\bibinfo
  {journal} {European Physical Journal --- Data Science}\ }\textbf {\bibinfo
  {volume} {10}},\ \bibinfo {eid} {51} (\bibinfo {year} {2021})}\BibitemShut
  {NoStop}%
\bibitem [{\citenamefont {{World Health Organization}}(2023)}]{who-infodemic}%
  \BibitemOpen
  \bibfield  {author} {\bibinfo {author} {\bibnamefont {{World Health
  Organization}}},\ }\href@noop {} {\enquote {\bibinfo {title} {Infodemic},}\
  }\bibinfo {howpublished} {Available at
  \url{https://www.who.int/health-topics/infodemic\#tab=tab_1}} (\bibinfo
  {year} {2023}),\ \bibinfo {note} {last accessed 16 March 2023}\BibitemShut
  {NoStop}%
\bibitem [{\citenamefont {Calleja}\ \emph {et~al.}(2021)\citenamefont {Calleja}
  \emph {et~al.}}]{infodemic-agenda}%
  \BibitemOpen
  \bibfield  {author} {\bibinfo {author} {\bibfnamefont {N.}~\bibnamefont
  {Calleja}} \emph {et~al.},\ }\bibfield  {title} {\enquote {\bibinfo {title}
  {A public health research agenda for managing infodemics: {M}ethods and
  results of the first {WHO} infodemiology conference},}\ }\href@noop {}
  {\bibfield  {journal} {\bibinfo  {journal} {JMIR Infodemiology}\ }\textbf
  {\bibinfo {volume} {1}},\ \bibinfo {eid} {e30979} (\bibinfo {year}
  {2021})}\BibitemShut {NoStop}%
\bibitem [{\citenamefont {Zarocostas}(2020)}]{zarocostas2020}%
  \BibitemOpen
  \bibfield  {author} {\bibinfo {author} {\bibfnamefont {J.}~\bibnamefont
  {Zarocostas}},\ }\bibfield  {title} {\enquote {\bibinfo {title} {How to fight
  an infodemic},}\ }\href@noop {} {\bibfield  {journal} {\bibinfo  {journal}
  {The Lancet}\ }\textbf {\bibinfo {volume} {395}},\ \bibinfo {eid} {676}
  (\bibinfo {year} {2020})}\BibitemShut {NoStop}%
\bibitem [{\citenamefont {Cinelli}\ \emph {et~al.}(2020)\citenamefont
  {Cinelli}, \citenamefont {Quattrociocchi}, \citenamefont {Galeazzi},
  \citenamefont {Valensise}, \citenamefont {Brugnoli}, \citenamefont {Schmidt},
  \citenamefont {Zollo},\ and\ \citenamefont {Scala}}]{cinelli2020}%
  \BibitemOpen
  \bibfield  {author} {\bibinfo {author} {\bibfnamefont {M.}~\bibnamefont
  {Cinelli}}, \bibinfo {author} {\bibfnamefont {W.}~\bibnamefont
  {Quattrociocchi}}, \bibinfo {author} {\bibfnamefont {A.}~\bibnamefont
  {Galeazzi}}, \bibinfo {author} {\bibfnamefont {C.~M.}\ \bibnamefont
  {Valensise}}, \bibinfo {author} {\bibfnamefont {E.}~\bibnamefont {Brugnoli}},
  \bibinfo {author} {\bibfnamefont {A.~L.}\ \bibnamefont {Schmidt}}, \bibinfo
  {author} {\bibfnamefont {P.~Z.~F.}\ \bibnamefont {Zollo}}, \ and\ \bibinfo
  {author} {\bibfnamefont {A.}~\bibnamefont {Scala}},\ }\bibfield  {title}
  {\enquote {\bibinfo {title} {The {COVID-19} social media infodemic},}\
  }\href@noop {} {\bibfield  {journal} {\bibinfo  {journal} {Scientific
  Reports}\ }\textbf {\bibinfo {volume} {10}},\ \bibinfo {eid} {16598}
  (\bibinfo {year} {2020})}\BibitemShut {NoStop}%
\bibitem [{\citenamefont {Gallotti}\ \emph {et~al.}(2020)\citenamefont
  {Gallotti}, \citenamefont {Valle}, \citenamefont {Castaldo}, \citenamefont
  {Sacco},\ and\ \citenamefont {Domenico}}]{gallotti2020}%
  \BibitemOpen
  \bibfield  {author} {\bibinfo {author} {\bibfnamefont {R.}~\bibnamefont
  {Gallotti}}, \bibinfo {author} {\bibfnamefont {F.}~\bibnamefont {Valle}},
  \bibinfo {author} {\bibfnamefont {N.}~\bibnamefont {Castaldo}}, \bibinfo
  {author} {\bibfnamefont {P.}~\bibnamefont {Sacco}}, \ and\ \bibinfo {author}
  {\bibfnamefont {M.~D.}\ \bibnamefont {Domenico}},\ }\bibfield  {title}
  {\enquote {\bibinfo {title} {Assessing the risks of `infodemics' in response
  to {COVID-19} epidemics},}\ }\href@noop {} {\bibfield  {journal} {\bibinfo
  {journal} {Nature Human Behaviour}\ }\textbf {\bibinfo {volume} {4}},\
  \bibinfo {pages} {1285--1293} (\bibinfo {year} {2020})}\BibitemShut {NoStop}%
\bibitem [{\citenamefont {Yang}\ \emph {et~al.}(2021)\citenamefont {Yang},
  \citenamefont {Pierri}, \citenamefont {Hui}, \citenamefont {Axelrod},
  \citenamefont {Torres-Lugo}, \citenamefont {Bryden},\ and\ \citenamefont
  {Menczer}}]{yang2021}%
  \BibitemOpen
  \bibfield  {author} {\bibinfo {author} {\bibfnamefont {K.-C.}\ \bibnamefont
  {Yang}}, \bibinfo {author} {\bibfnamefont {F.}~\bibnamefont {Pierri}},
  \bibinfo {author} {\bibfnamefont {P.-M.}\ \bibnamefont {Hui}}, \bibinfo
  {author} {\bibfnamefont {D.}~\bibnamefont {Axelrod}}, \bibinfo {author}
  {\bibfnamefont {C.}~\bibnamefont {Torres-Lugo}}, \bibinfo {author}
  {\bibfnamefont {J.}~\bibnamefont {Bryden}}, \ and\ \bibinfo {author}
  {\bibfnamefont {F.}~\bibnamefont {Menczer}},\ }\bibfield  {title} {\enquote
  {\bibinfo {title} {The {COVID-19} infodemic: {T}witter versus {F}acebook},}\
  }\href@noop {} {\bibfield  {journal} {\bibinfo  {journal} {Big Data \&
  Society}\ }\textbf {\bibinfo {volume} {8}},\ \bibinfo {eid}
  {20539517211013861} (\bibinfo {year} {2021})}\BibitemShut {NoStop}%
\bibitem [{\citenamefont {Lazer}\ \emph {et~al.}(2018)\citenamefont {Lazer},
  \citenamefont {Baum}, \citenamefont {Benkler}, \citenamefont {Berinsky},
  \citenamefont {Greenhill}, \citenamefont {Menczer}, \citenamefont {Metzger},
  \citenamefont {Nyhan}, \citenamefont {Pennycook}, \citenamefont {Rothschild},
  \citenamefont {Schudson}, \citenamefont {Sloman}, \citenamefont {Sunstein},
  \citenamefont {Thorson}, \citenamefont {Watts},\ and\ \citenamefont
  {Zittrain}}]{lazer2018}%
  \BibitemOpen
  \bibfield  {author} {\bibinfo {author} {\bibfnamefont {D.~M.~J.}\
  \bibnamefont {Lazer}}, \bibinfo {author} {\bibfnamefont {M.~A.}\ \bibnamefont
  {Baum}}, \bibinfo {author} {\bibfnamefont {Y.}~\bibnamefont {Benkler}},
  \bibinfo {author} {\bibfnamefont {A.~J.}\ \bibnamefont {Berinsky}}, \bibinfo
  {author} {\bibfnamefont {K.~M.}\ \bibnamefont {Greenhill}}, \bibinfo {author}
  {\bibfnamefont {F.}~\bibnamefont {Menczer}}, \bibinfo {author} {\bibfnamefont
  {M.~J.}\ \bibnamefont {Metzger}}, \bibinfo {author} {\bibfnamefont
  {B.}~\bibnamefont {Nyhan}}, \bibinfo {author} {\bibfnamefont
  {G.}~\bibnamefont {Pennycook}}, \bibinfo {author} {\bibfnamefont
  {D.}~\bibnamefont {Rothschild}}, \bibinfo {author} {\bibfnamefont
  {M.}~\bibnamefont {Schudson}}, \bibinfo {author} {\bibfnamefont {S.~A.}\
  \bibnamefont {Sloman}}, \bibinfo {author} {\bibfnamefont {C.~R.}\
  \bibnamefont {Sunstein}}, \bibinfo {author} {\bibfnamefont {E.~A.}\
  \bibnamefont {Thorson}}, \bibinfo {author} {\bibfnamefont {D.~J.}\
  \bibnamefont {Watts}}, \ and\ \bibinfo {author} {\bibfnamefont {J.~L.}\
  \bibnamefont {Zittrain}},\ }\bibfield  {title} {\enquote {\bibinfo {title}
  {The science of fake news},}\ }\href@noop {} {\bibfield  {journal} {\bibinfo
  {journal} {Science}\ }\textbf {\bibinfo {volume} {369}},\ \bibinfo {pages}
  {1094--1096} (\bibinfo {year} {2018})}\BibitemShut {NoStop}%
\bibitem [{\citenamefont {Watts}, \citenamefont {Rothschild},\ and\
  \citenamefont {Mobius}(2021)}]{watts2021}%
  \BibitemOpen
  \bibfield  {author} {\bibinfo {author} {\bibfnamefont {D.~J.}\ \bibnamefont
  {Watts}}, \bibinfo {author} {\bibfnamefont {D.~M.}\ \bibnamefont
  {Rothschild}}, \ and\ \bibinfo {author} {\bibfnamefont {M.}~\bibnamefont
  {Mobius}},\ }\bibfield  {title} {\enquote {\bibinfo {title} {Measuring the
  news and its impact on democracy},}\ }\href@noop {} {\bibfield  {journal}
  {\bibinfo  {journal} {Proceedings of the National Academy of Sciences of the
  United States of America}\ }\textbf {\bibinfo {volume} {118}},\ \bibinfo
  {eid} {e1912443118} (\bibinfo {year} {2021})}\BibitemShut {NoStop}%
\bibitem [{\citenamefont {Rabb}\ \emph {et~al.}(2022)\citenamefont {Rabb},
  \citenamefont {Cowen}, \citenamefont {de~Ruiter},\ and\ \citenamefont
  {Scheutz}}]{rabb2022}%
  \BibitemOpen
  \bibfield  {author} {\bibinfo {author} {\bibfnamefont {N.}~\bibnamefont
  {Rabb}}, \bibinfo {author} {\bibfnamefont {L.}~\bibnamefont {Cowen}},
  \bibinfo {author} {\bibfnamefont {J.~P.}\ \bibnamefont {de~Ruiter}}, \ and\
  \bibinfo {author} {\bibfnamefont {M.}~\bibnamefont {Scheutz}},\ }\bibfield
  {title} {\enquote {\bibinfo {title} {Cognitive cascades: {H}ow to model (and
  potentially counter) the spread of fake news},}\ }\href@noop {} {\bibfield
  {journal} {\bibinfo  {journal} {PLoS ONE}\ }\textbf {\bibinfo {volume}
  {17}},\ \bibinfo {eid} {e0261811} (\bibinfo {year} {2022})}\BibitemShut
  {NoStop}%
\bibitem [{\citenamefont {Noorazar}\ \emph {et~al.}(2020)\citenamefont
  {Noorazar}, \citenamefont {Vixie}, \citenamefont {Talebanpour},\ and\
  \citenamefont {Hu}}]{noorazar2020classical}%
  \BibitemOpen
  \bibfield  {author} {\bibinfo {author} {\bibfnamefont {H.}~\bibnamefont
  {Noorazar}}, \bibinfo {author} {\bibfnamefont {K.~R.}\ \bibnamefont {Vixie}},
  \bibinfo {author} {\bibfnamefont {A.}~\bibnamefont {Talebanpour}}, \ and\
  \bibinfo {author} {\bibfnamefont {Y.}~\bibnamefont {Hu}},\ }\bibfield
  {title} {\enquote {\bibinfo {title} {From classical to modern opinion
  dynamics},}\ }\href@noop {} {\bibfield  {journal} {\bibinfo  {journal}
  {International Journal of Modern Physics C}\ }\textbf {\bibinfo {volume}
  {31}},\ \bibinfo {eid} {2050101} (\bibinfo {year} {2020})}\BibitemShut
  {NoStop}%
\bibitem [{\citenamefont {Noorazar}(2020)}]{noorazar2020recent}%
  \BibitemOpen
  \bibfield  {author} {\bibinfo {author} {\bibfnamefont {H.}~\bibnamefont
  {Noorazar}},\ }\bibfield  {title} {\enquote {\bibinfo {title} {Recent
  advances in opinion propagation dynamics: {A} 2020 survey},}\ }\href@noop {}
  {\bibfield  {journal} {\bibinfo  {journal} {The European Physical Journal
  Plus}\ }\textbf {\bibinfo {volume} {135}},\ \bibinfo {eid} {521} (\bibinfo
  {year} {2020})}\BibitemShut {NoStop}%
\bibitem [{\citenamefont {Peralta}, \citenamefont {Kert\'{e}sz},\ and\
  \citenamefont {I{\~n}iguez}(2022)}]{peralta2022}%
  \BibitemOpen
  \bibfield  {author} {\bibinfo {author} {\bibfnamefont {A.~F.}\ \bibnamefont
  {Peralta}}, \bibinfo {author} {\bibfnamefont {J.}~\bibnamefont
  {Kert\'{e}sz}}, \ and\ \bibinfo {author} {\bibfnamefont {G.}~\bibnamefont
  {I{\~n}iguez}},\ }\bibfield  {title} {\enquote {\bibinfo {title} {Opinion
  dynamics in social networks: {F}rom models to data},}\ }\href@noop {}
  {\bibfield  {journal} {\bibinfo  {journal} {arXiv preprint arXiv:2201.01322}\
  } (\bibinfo {year} {2022})}\BibitemShut {NoStop}%
\bibitem [{\citenamefont {Lorenz}(2007)}]{lorenz2007continuous}%
  \BibitemOpen
  \bibfield  {author} {\bibinfo {author} {\bibfnamefont {J.}~\bibnamefont
  {Lorenz}},\ }\bibfield  {title} {\enquote {\bibinfo {title} {Continuous
  opinion dynamics under bounded confidence: {A} survey},}\ }\href@noop {}
  {\bibfield  {journal} {\bibinfo  {journal} {International Journal of Modern
  Physics C}\ }\textbf {\bibinfo {volume} {18}},\ \bibinfo {pages} {1819--1838}
  (\bibinfo {year} {2007})}\BibitemShut {NoStop}%
\bibitem [{\citenamefont {Meng}, \citenamefont {Van~Gorder},\ and\
  \citenamefont {Porter}(2018)}]{meng2018opinion}%
  \BibitemOpen
  \bibfield  {author} {\bibinfo {author} {\bibfnamefont {X.~F.}\ \bibnamefont
  {Meng}}, \bibinfo {author} {\bibfnamefont {R.~A.}\ \bibnamefont
  {Van~Gorder}}, \ and\ \bibinfo {author} {\bibfnamefont {M.~A.}\ \bibnamefont
  {Porter}},\ }\bibfield  {title} {\enquote {\bibinfo {title} {Opinion
  formation and distribution in a bounded-confidence model on various
  networks},}\ }\href@noop {} {\bibfield  {journal} {\bibinfo  {journal}
  {Physical Review E}\ }\textbf {\bibinfo {volume} {97}},\ \bibinfo {eid}
  {022312} (\bibinfo {year} {2018})}\BibitemShut {NoStop}%
\bibitem [{\citenamefont {Bernardo}\ \emph {et~al.}(2024)\citenamefont
  {Bernardo}, \citenamefont {Altafini}, \citenamefont {Proskurnikov},\ and\
  \citenamefont {Vasca}}]{bernardo2024}%
  \BibitemOpen
  \bibfield  {author} {\bibinfo {author} {\bibfnamefont {C.}~\bibnamefont
  {Bernardo}}, \bibinfo {author} {\bibfnamefont {C.}~\bibnamefont {Altafini}},
  \bibinfo {author} {\bibfnamefont {A.}~\bibnamefont {Proskurnikov}}, \ and\
  \bibinfo {author} {\bibfnamefont {F.}~\bibnamefont {Vasca}},\ }\bibfield
  {title} {\enquote {\bibinfo {title} {Bounded confidence opinion dynamics: {A}
  survey},}\ }\href@noop {} {\bibfield  {journal} {\bibinfo  {journal}
  {Automatica}\ }\textbf {\bibinfo {volume} {159}},\ \bibinfo {eid} {111302}
  (\bibinfo {year} {2024})}\BibitemShut {NoStop}%
\bibitem [{\citenamefont {Brooks}\ and\ \citenamefont
  {Porter}(2020)}]{brooks2020model}%
  \BibitemOpen
  \bibfield  {author} {\bibinfo {author} {\bibfnamefont {H.~Z.}\ \bibnamefont
  {Brooks}}\ and\ \bibinfo {author} {\bibfnamefont {M.~A.}\ \bibnamefont
  {Porter}},\ }\bibfield  {title} {\enquote {\bibinfo {title} {A model for the
  influence of media on the ideology of content in online social networks},}\
  }\href@noop {} {\bibfield  {journal} {\bibinfo  {journal} {Physical Review
  Research}\ }\textbf {\bibinfo {volume} {2}},\ \bibinfo {eid} {023041}
  (\bibinfo {year} {2020})}\BibitemShut {NoStop}%
\bibitem [{\citenamefont {Hickok}\ \emph {et~al.}(2022)\citenamefont {Hickok},
  \citenamefont {Kureh}, \citenamefont {Brooks}, \citenamefont {Feng},\ and\
  \citenamefont {Porter}}]{hickok2022bounded}%
  \BibitemOpen
  \bibfield  {author} {\bibinfo {author} {\bibfnamefont {A.}~\bibnamefont
  {Hickok}}, \bibinfo {author} {\bibfnamefont {Y.}~\bibnamefont {Kureh}},
  \bibinfo {author} {\bibfnamefont {H.~Z.}\ \bibnamefont {Brooks}}, \bibinfo
  {author} {\bibfnamefont {M.}~\bibnamefont {Feng}}, \ and\ \bibinfo {author}
  {\bibfnamefont {M.~A.}\ \bibnamefont {Porter}},\ }\bibfield  {title}
  {\enquote {\bibinfo {title} {A bounded-confidence model of opinion dynamics
  on hypergraphs},}\ }\href@noop {} {\bibfield  {journal} {\bibinfo  {journal}
  {SIAM Journal on Applied Dynamical Systems}\ }\textbf {\bibinfo {volume}
  {21}},\ \bibinfo {pages} {1--32} (\bibinfo {year} {2022})}\BibitemShut
  {NoStop}%
\bibitem [{\citenamefont {Wang}(2022)}]{wang2022opinion}%
  \BibitemOpen
  \bibfield  {author} {\bibinfo {author} {\bibfnamefont {C.}~\bibnamefont
  {Wang}},\ }\bibfield  {title} {\enquote {\bibinfo {title} {Opinion dynamics
  with higher-order bounded confidence},}\ }\href@noop {} {\bibfield  {journal}
  {\bibinfo  {journal} {Entropy}\ }\textbf {\bibinfo {volume} {24}},\ \bibinfo
  {pages} {1300} (\bibinfo {year} {2022})}\BibitemShut {NoStop}%
\bibitem [{\citenamefont {Goddard}\ \emph {et~al.}(2022)\citenamefont
  {Goddard}, \citenamefont {Gooding}, \citenamefont {Short},\ and\
  \citenamefont {Pavliotis}}]{goddard2022noisy}%
  \BibitemOpen
  \bibfield  {author} {\bibinfo {author} {\bibfnamefont {B.~D.}\ \bibnamefont
  {Goddard}}, \bibinfo {author} {\bibfnamefont {B.}~\bibnamefont {Gooding}},
  \bibinfo {author} {\bibfnamefont {H.}~\bibnamefont {Short}}, \ and\ \bibinfo
  {author} {\bibfnamefont {G.~A.}\ \bibnamefont {Pavliotis}},\ }\bibfield
  {title} {\enquote {\bibinfo {title} {Noisy bounded confidence models for
  opinion dynamics: {T}he effect of boundary conditions on phase
  transitions},}\ }\href@noop {} {\bibfield  {journal} {\bibinfo  {journal}
  {IMA Journal of Applied Mathematics}\ }\textbf {\bibinfo {volume} {87}},\
  \bibinfo {pages} {80--110} (\bibinfo {year} {2022})}\BibitemShut {NoStop}%
\bibitem [{\citenamefont {Bernardo}, \citenamefont {Altafini},\ and\
  \citenamefont {Vasca}(2022)}]{bernardo2022finite}%
  \BibitemOpen
  \bibfield  {author} {\bibinfo {author} {\bibfnamefont {C.}~\bibnamefont
  {Bernardo}}, \bibinfo {author} {\bibfnamefont {C.}~\bibnamefont {Altafini}},
  \ and\ \bibinfo {author} {\bibfnamefont {F.}~\bibnamefont {Vasca}},\
  }\bibfield  {title} {\enquote {\bibinfo {title} {Finite-time convergence of
  opinion dynamics in homogeneous asymmetric bounded confidence models},}\
  }\href@noop {} {\bibfield  {journal} {\bibinfo  {journal} {European Journal
  of Control}\ }\textbf {\bibinfo {volume} {68}},\ \bibinfo {eid} {100674}
  (\bibinfo {year} {2022})}\BibitemShut {NoStop}%
\bibitem [{\citenamefont {Schawe}\ and\ \citenamefont
  {Hern{\'a}ndez}(2020)}]{schawe2020collective}%
  \BibitemOpen
  \bibfield  {author} {\bibinfo {author} {\bibfnamefont {H.}~\bibnamefont
  {Schawe}}\ and\ \bibinfo {author} {\bibfnamefont {L.}~\bibnamefont
  {Hern{\'a}ndez}},\ }\bibfield  {title} {\enquote {\bibinfo {title}
  {Collective effects of the cost of opinion change},}\ }\href@noop {}
  {\bibfield  {journal} {\bibinfo  {journal} {Scientific Reports}\ }\textbf
  {\bibinfo {volume} {10}},\ \bibinfo {pages} {13825} (\bibinfo {year}
  {2020})}\BibitemShut {NoStop}%
\bibitem [{\citenamefont {Li}\ and\ \citenamefont {Porter}(2023)}]{grace2023}%
  \BibitemOpen
  \bibfield  {author} {\bibinfo {author} {\bibfnamefont {G.~J.}\ \bibnamefont
  {Li}}\ and\ \bibinfo {author} {\bibfnamefont {M.~A.}\ \bibnamefont
  {Porter}},\ }\bibfield  {title} {\enquote {\bibinfo {title}
  {Bounded-confidence model of opinion dynamics with heterogeneous
  node-activity levels},}\ }\href@noop {} {\bibfield  {journal} {\bibinfo
  {journal} {Physical Review Research}\ }\textbf {\bibinfo {volume} {5}},\
  \bibinfo {eid} {023179} (\bibinfo {year} {2023})}\BibitemShut {NoStop}%
\bibitem [{\citenamefont {Brooks}, \citenamefont {Chodrow},\ and\ \citenamefont
  {Porter}(2024)}]{brooks2022emergence}%
  \BibitemOpen
  \bibfield  {author} {\bibinfo {author} {\bibfnamefont {H.~Z.}\ \bibnamefont
  {Brooks}}, \bibinfo {author} {\bibfnamefont {P.~S.}\ \bibnamefont {Chodrow}},
  \ and\ \bibinfo {author} {\bibfnamefont {M.~A.}\ \bibnamefont {Porter}},\
  }\bibfield  {title} {\enquote {\bibinfo {title} {Emergence of polarization in
  a sigmoidal bounded-confidence model of opinion dynamics},}\ }\href@noop {}
  {\bibfield  {journal} {\bibinfo  {journal} {SIAM Journal on Applied Dynamical
  Systems}\ }\textbf {\bibinfo {volume} {23}},\ \bibinfo {pages} {1442--1470}
  (\bibinfo {year} {2024})}\BibitemShut {NoStop}%
\bibitem [{\citenamefont {Kann}\ and\ \citenamefont
  {Feng}(2023)}]{kann2023repulsive}%
  \BibitemOpen
  \bibfield  {author} {\bibinfo {author} {\bibfnamefont {C.}~\bibnamefont
  {Kann}}\ and\ \bibinfo {author} {\bibfnamefont {M.}~\bibnamefont {Feng}},\
  }\bibfield  {title} {\enquote {\bibinfo {title} {Repulsive bounded-confidence
  model of opinion dynamics in polarized communities},}\ }\href@noop {}
  {\bibfield  {journal} {\bibinfo  {journal} {arXiv preprint arXiv:2301.02210}\
  } (\bibinfo {year} {2023})}\BibitemShut {NoStop}%
\bibitem [{\citenamefont {Kan}, \citenamefont {Feng},\ and\ \citenamefont
  {Porter}(2023)}]{kan2023adaptive}%
  \BibitemOpen
  \bibfield  {author} {\bibinfo {author} {\bibfnamefont {U.}~\bibnamefont
  {Kan}}, \bibinfo {author} {\bibfnamefont {M.}~\bibnamefont {Feng}}, \ and\
  \bibinfo {author} {\bibfnamefont {M.~A.}\ \bibnamefont {Porter}},\ }\bibfield
   {title} {\enquote {\bibinfo {title} {An adaptive bounded-confidence model of
  opinion dynamics on networks},}\ }\href@noop {} {\bibfield  {journal}
  {\bibinfo  {journal} {Journal of Complex Networks}\ }\textbf {\bibinfo
  {volume} {11}},\ \bibinfo {eid} {cnac055} (\bibinfo {year}
  {2023})}\BibitemShut {NoStop}%
\bibitem [{\citenamefont {Li}, \citenamefont {Motsch},\ and\ \citenamefont
  {Weber}(2020)}]{li2020bounded}%
  \BibitemOpen
  \bibfield  {author} {\bibinfo {author} {\bibfnamefont {G.~L.}\ \bibnamefont
  {Li}}, \bibinfo {author} {\bibfnamefont {S.}~\bibnamefont {Motsch}}, \ and\
  \bibinfo {author} {\bibfnamefont {D.}~\bibnamefont {Weber}},\ }\bibfield
  {title} {\enquote {\bibinfo {title} {Bounded confidence dynamics and graph
  control: {E}nforcing consensus},}\ }\href@noop {} {\bibfield  {journal}
  {\bibinfo  {journal} {Networks and Heterogeneous Media}\ }\textbf {\bibinfo
  {volume} {15}},\ \bibinfo {pages} {489--517} (\bibinfo {year}
  {2020})}\BibitemShut {NoStop}%
\bibitem [{\citenamefont {Li}, \citenamefont {Luo},\ and\ \citenamefont
  {Porter}(2024)}]{grace-jerry2023}%
  \BibitemOpen
  \bibfield  {author} {\bibinfo {author} {\bibfnamefont {G.~J.}\ \bibnamefont
  {Li}}, \bibinfo {author} {\bibfnamefont {J.}~\bibnamefont {Luo}}, \ and\
  \bibinfo {author} {\bibfnamefont {M.~A.}\ \bibnamefont {Porter}},\ }\bibfield
   {title} {\enquote {\bibinfo {title} {Bounded-confidence models of opinion
  dynamics with adaptive confidence bounds},}\ }\href@noop {} {\bibfield
  {journal} {\bibinfo  {journal} {arXiv preprint arXiv:2303.07563 (SIAM Journal
  on Applied Dynamical Systems, in press)}\ } (\bibinfo {year}
  {2024})}\BibitemShut {NoStop}%
\bibitem [{\citenamefont {Porter}\ and\ \citenamefont
  {Gleeson}(2016)}]{porter2016}%
  \BibitemOpen
  \bibfield  {author} {\bibinfo {author} {\bibfnamefont {M.~A.}\ \bibnamefont
  {Porter}}\ and\ \bibinfo {author} {\bibfnamefont {J.~P.}\ \bibnamefont
  {Gleeson}},\ }\href@noop {} {\emph {\bibinfo {title} {Dynamical systems on
  networks: {A} tutorial}}},\ \bibinfo {series} {Frontiers in Applied Dynamical
  Systems: Reviews and Tutorials}, Vol.~\bibinfo {volume} {4}\ (\bibinfo
  {publisher} {Springer},\ \bibinfo {address} {Cham, Switzerland},\ \bibinfo
  {year} {2016})\BibitemShut {NoStop}%
\bibitem [{\citenamefont {Sahimi}(2023)}]{sahimi2023}%
  \BibitemOpen
  \bibfield  {author} {\bibinfo {author} {\bibfnamefont {M.}~\bibnamefont
  {Sahimi}},\ }\href@noop {} {\emph {\bibinfo {title} {Applications of
  Percolation Theory}}},\ \bibinfo {edition} {2nd}\ ed.\ (\bibinfo  {publisher}
  {Springer},\ \bibinfo {address} {Heidelberg, Germany},\ \bibinfo {year}
  {2023})\BibitemShut {NoStop}%
\bibitem [{\citenamefont {Malthe-S{\o}renssen}(2024)}]{percolation2024}%
  \BibitemOpen
  \bibfield  {author} {\bibinfo {author} {\bibfnamefont {A.}~\bibnamefont
  {Malthe-S{\o}renssen}},\ }\href@noop {} {\emph {\bibinfo {title} {Percolation
  Theory Using Python}}}\ (\bibinfo  {publisher} {Springer},\ \bibinfo
  {address} {Cham, Switzerland},\ \bibinfo {year} {2024})\BibitemShut {NoStop}%
\bibitem [{\citenamefont {Lehmann}\ and\ \citenamefont
  {Ahn}(2018)}]{lehmann2018}%
  \BibitemOpen
  \bibinfo {editor} {\bibfnamefont {S.}~\bibnamefont {Lehmann}}\ and\ \bibinfo
  {editor} {\bibfnamefont {Y.-Y.}\ \bibnamefont {Ahn}},\ eds.,\ \href@noop {}
  {\emph {\bibinfo {title} {Complex Spreading Phenomena in Social Systems:
  Influence and Contagion in Real-World Social Networks}}}\ (\bibinfo
  {publisher} {Springer},\ \bibinfo {address} {Cham, Switzerland},\ \bibinfo
  {year} {2018})\BibitemShut {NoStop}%
\bibitem [{\citenamefont {Pastor-Satorras}\ \emph {et~al.}(2015)\citenamefont
  {Pastor-Satorras}, \citenamefont {Castellano}, \citenamefont {Van~Mieghem},\
  and\ \citenamefont {Vespignani}}]{pastor2015}%
  \BibitemOpen
  \bibfield  {author} {\bibinfo {author} {\bibfnamefont {R.}~\bibnamefont
  {Pastor-Satorras}}, \bibinfo {author} {\bibfnamefont {C.}~\bibnamefont
  {Castellano}}, \bibinfo {author} {\bibfnamefont {P.}~\bibnamefont
  {Van~Mieghem}}, \ and\ \bibinfo {author} {\bibfnamefont {A.}~\bibnamefont
  {Vespignani}},\ }\bibfield  {title} {\enquote {\bibinfo {title} {Epidemic
  processes in complex networks},}\ }\href@noop {} {\bibfield  {journal}
  {\bibinfo  {journal} {Reviews of Modern Physics}\ }\textbf {\bibinfo {volume}
  {87}},\ \bibinfo {pages} {925--979} (\bibinfo {year} {2015})}\BibitemShut
  {NoStop}%
\bibitem [{\citenamefont {Bedson}\ \emph {et~al.}(2021)\citenamefont {Bedson},
  \citenamefont {Skrip}, \citenamefont {Pedi}, \citenamefont {Abramowitz},
  \citenamefont {Carter}, \citenamefont {Jalloh}, \citenamefont {Funk},
  \citenamefont {Gobat}, \citenamefont {Giles-Vernick}, \citenamefont
  {Chowell}, \citenamefont {de~Almeida}, \citenamefont {Elessawi},
  \citenamefont {Scarpino}, \citenamefont {Hammond}, \citenamefont {Briand},
  \citenamefont {Epstein}, \citenamefont {H{\'e}bert-Dufresne},\ and\
  \citenamefont {Althouse}}]{bedson2021}%
  \BibitemOpen
  \bibfield  {author} {\bibinfo {author} {\bibfnamefont {J.}~\bibnamefont
  {Bedson}}, \bibinfo {author} {\bibfnamefont {L.~A.}\ \bibnamefont {Skrip}},
  \bibinfo {author} {\bibfnamefont {D.}~\bibnamefont {Pedi}}, \bibinfo {author}
  {\bibfnamefont {S.}~\bibnamefont {Abramowitz}}, \bibinfo {author}
  {\bibfnamefont {S.}~\bibnamefont {Carter}}, \bibinfo {author} {\bibfnamefont
  {M.~F.}\ \bibnamefont {Jalloh}}, \bibinfo {author} {\bibfnamefont
  {S.}~\bibnamefont {Funk}}, \bibinfo {author} {\bibfnamefont {N.}~\bibnamefont
  {Gobat}}, \bibinfo {author} {\bibfnamefont {T.}~\bibnamefont
  {Giles-Vernick}}, \bibinfo {author} {\bibfnamefont {G.}~\bibnamefont
  {Chowell}}, \bibinfo {author} {\bibfnamefont {J.~R.}\ \bibnamefont
  {de~Almeida}}, \bibinfo {author} {\bibfnamefont {R.}~\bibnamefont
  {Elessawi}}, \bibinfo {author} {\bibfnamefont {S.~V.}\ \bibnamefont
  {Scarpino}}, \bibinfo {author} {\bibfnamefont {R.~A.}\ \bibnamefont
  {Hammond}}, \bibinfo {author} {\bibfnamefont {S.}~\bibnamefont {Briand}},
  \bibinfo {author} {\bibfnamefont {J.~M.}\ \bibnamefont {Epstein}}, \bibinfo
  {author} {\bibfnamefont {L.}~\bibnamefont {H{\'e}bert-Dufresne}}, \ and\
  \bibinfo {author} {\bibfnamefont {B.~M.}\ \bibnamefont {Althouse}},\
  }\bibfield  {title} {\enquote {\bibinfo {title} {A review and agenda for
  integrated disease models including social and behavioural factors},}\
  }\href@noop {} {\bibfield  {journal} {\bibinfo  {journal} {Nature Human
  Behaviour}\ }\textbf {\bibinfo {volume} {5}},\ \bibinfo {pages} {834--846}
  (\bibinfo {year} {2021})}\BibitemShut {NoStop}%
\bibitem [{\citenamefont {Weng}, \citenamefont {Menczer},\ and\ \citenamefont
  {Ahn}(2013)}]{weng2013}%
  \BibitemOpen
  \bibfield  {author} {\bibinfo {author} {\bibfnamefont {L.}~\bibnamefont
  {Weng}}, \bibinfo {author} {\bibfnamefont {F.}~\bibnamefont {Menczer}}, \
  and\ \bibinfo {author} {\bibfnamefont {Y.-Y.}\ \bibnamefont {Ahn}},\
  }\bibfield  {title} {\enquote {\bibinfo {title} {Virality prediction and
  community structure in social networks},}\ }\href@noop {} {\bibfield
  {journal} {\bibinfo  {journal} {Scientific Reports}\ }\textbf {\bibinfo
  {volume} {3}},\ \bibinfo {eid} {2522} (\bibinfo {year} {2013})}\BibitemShut
  {NoStop}%
\bibitem [{\citenamefont {Lerman}(2016)}]{lerman2016}%
  \BibitemOpen
  \bibfield  {author} {\bibinfo {author} {\bibfnamefont {K.}~\bibnamefont
  {Lerman}},\ }\bibfield  {title} {\enquote {\bibinfo {title} {Information is
  not a virus, and other consequences of human cognitive limits},}\ }\href@noop
  {} {\bibfield  {journal} {\bibinfo  {journal} {Future Internet}\ }\textbf
  {\bibinfo {volume} {8}},\ \bibinfo {eid} {21} (\bibinfo {year}
  {2016})}\BibitemShut {NoStop}%
\bibitem [{\citenamefont {Hui}\ \emph {et~al.}(2018)\citenamefont {Hui},
  \citenamefont {Weng}, \citenamefont {Sahami~Shirazi}, \citenamefont {Ahn},\
  and\ \citenamefont {Menczer}}]{hui2018}%
  \BibitemOpen
  \bibfield  {author} {\bibinfo {author} {\bibfnamefont {P.-M.}\ \bibnamefont
  {Hui}}, \bibinfo {author} {\bibfnamefont {L.}~\bibnamefont {Weng}}, \bibinfo
  {author} {\bibfnamefont {A.}~\bibnamefont {Sahami~Shirazi}}, \bibinfo
  {author} {\bibfnamefont {Y.-Y.}\ \bibnamefont {Ahn}}, \ and\ \bibinfo
  {author} {\bibfnamefont {F.}~\bibnamefont {Menczer}},\ }\enquote {\bibinfo
  {title} {Scalable detection of viral memes from diffusion patterns},}\ in\
  \href@noop {} {\emph {\bibinfo {booktitle} {Complex Spreading Phenomena in
  Social Systems: Influence and Contagion in Real-World Social Networks}}},\
  \bibinfo {editor} {edited by\ \bibinfo {editor} {\bibfnamefont
  {S.}~\bibnamefont {Lehmann}}\ and\ \bibinfo {editor} {\bibfnamefont {Y.-Y.}\
  \bibnamefont {Ahn}}}\ (\bibinfo  {publisher} {Springer},\ \bibinfo {address}
  {Cham, Switzerland},\ \bibinfo {year} {2018})\ pp.\ \bibinfo {pages}
  {197--211}\BibitemShut {NoStop}%
\bibitem [{\citenamefont {Olsson}\ and\ \citenamefont
  {Galesic}(2024)}]{olsson2023}%
  \BibitemOpen
  \bibfield  {author} {\bibinfo {author} {\bibfnamefont {H.}~\bibnamefont
  {Olsson}}\ and\ \bibinfo {author} {\bibfnamefont {M.}~\bibnamefont
  {Galesic}},\ }\bibfield  {title} {\enquote {\bibinfo {title} {Analogies for
  modeling belief dynamics},}\ }\href@noop {} {\bibfield  {journal} {\bibinfo
  {journal} {Trends in Cognitive Sciences}\ } (\bibinfo {year} {2024})},\
  \bibinfo {note} {available at
  \url{https://doi.org/10.1016/j.tics.2024.07.001}}\BibitemShut {NoStop}%
\bibitem [{\citenamefont {Cencetti}\ \emph {et~al.}(2023)\citenamefont
  {Cencetti}, \citenamefont {Contreras}, \citenamefont {Mancastroppa},\ and\
  \citenamefont {Barrat}}]{cencetti2023}%
  \BibitemOpen
  \bibfield  {author} {\bibinfo {author} {\bibfnamefont {G.}~\bibnamefont
  {Cencetti}}, \bibinfo {author} {\bibfnamefont {D.~A.}\ \bibnamefont
  {Contreras}}, \bibinfo {author} {\bibfnamefont {M.}~\bibnamefont
  {Mancastroppa}}, \ and\ \bibinfo {author} {\bibfnamefont {A.}~\bibnamefont
  {Barrat}},\ }\bibfield  {title} {\enquote {\bibinfo {title} {Distinguishing
  simple and complex contagion processes on networks},}\ }\href@noop {}
  {\bibfield  {journal} {\bibinfo  {journal} {Physical Review Letters}\
  }\textbf {\bibinfo {volume} {130}},\ \bibinfo {pages} {247401} (\bibinfo
  {year} {2023})}\BibitemShut {NoStop}%
\bibitem [{\citenamefont {St-Onge}, \citenamefont {Hébert-Dufresne},\ and\
  \citenamefont {Allard}(2024)}]{st-onge2024}%
  \BibitemOpen
  \bibfield  {author} {\bibinfo {author} {\bibfnamefont {G.}~\bibnamefont
  {St-Onge}}, \bibinfo {author} {\bibfnamefont {L.}~\bibnamefont
  {Hébert-Dufresne}}, \ and\ \bibinfo {author} {\bibfnamefont
  {A.}~\bibnamefont {Allard}},\ }\bibfield  {title} {\enquote {\bibinfo {title}
  {Nonlinear bias toward complex contagion in uncertain transmission
  settings},}\ }\href@noop {} {\bibfield  {journal} {\bibinfo  {journal}
  {Proceedings of the National Academy of Sciences of the United States of
  America}\ }\textbf {\bibinfo {volume} {121}},\ \bibinfo {eid} {e2312202121}
  (\bibinfo {year} {2024})}\BibitemShut {NoStop}%
\bibitem [{\citenamefont {Contreras}, \citenamefont {Cencetti},\ and\
  \citenamefont {Barrat}(2024)}]{contreras2024}%
  \BibitemOpen
  \bibfield  {author} {\bibinfo {author} {\bibfnamefont {D.~A.}\ \bibnamefont
  {Contreras}}, \bibinfo {author} {\bibfnamefont {G.}~\bibnamefont {Cencetti}},
  \ and\ \bibinfo {author} {\bibfnamefont {A.}~\bibnamefont {Barrat}},\
  }\bibfield  {title} {\enquote {\bibinfo {title} {Infection patterns in simple
  and complex contagion processes on networks},}\ }\href@noop {} {\bibfield
  {journal} {\bibinfo  {journal} {PLoS Computational Biology}\ }\textbf
  {\bibinfo {volume} {20}},\ \bibinfo {pages} {e1012206} (\bibinfo {year}
  {2024})}\BibitemShut {NoStop}%
\bibitem [{\citenamefont {No\"el}\ \emph {et~al.}(2012)\citenamefont {No\"el},
  \citenamefont {Allard}, \citenamefont {H\'ebert-Dufresne}, \citenamefont
  {Marceau},\ and\ \citenamefont {Dub\'e}}]{noel2012}%
  \BibitemOpen
  \bibfield  {author} {\bibinfo {author} {\bibfnamefont {P.-A.}\ \bibnamefont
  {No\"el}}, \bibinfo {author} {\bibfnamefont {A.}~\bibnamefont {Allard}},
  \bibinfo {author} {\bibfnamefont {L.}~\bibnamefont {H\'ebert-Dufresne}},
  \bibinfo {author} {\bibfnamefont {V.}~\bibnamefont {Marceau}}, \ and\
  \bibinfo {author} {\bibfnamefont {L.~J.}\ \bibnamefont {Dub\'e}},\ }\bibfield
   {title} {\enquote {\bibinfo {title} {Propagation on networks: {A}n exact
  alternative perspective},}\ }\href@noop {} {\bibfield  {journal} {\bibinfo
  {journal} {Physical Review E}\ }\textbf {\bibinfo {volume} {85}},\ \bibinfo
  {eid} {031118} (\bibinfo {year} {2012})}\BibitemShut {NoStop}%
\bibitem [{\citenamefont {Gomez~Rodriguez}, \citenamefont {Leskovec},\ and\
  \citenamefont {Krause}(2010)}]{gomez2010}%
  \BibitemOpen
  \bibfield  {author} {\bibinfo {author} {\bibfnamefont {M.}~\bibnamefont
  {Gomez~Rodriguez}}, \bibinfo {author} {\bibfnamefont {J.}~\bibnamefont
  {Leskovec}}, \ and\ \bibinfo {author} {\bibfnamefont {A.}~\bibnamefont
  {Krause}},\ }\bibfield  {title} {\enquote {\bibinfo {title} {Inferring
  networks of diffusion and influence},}\ }in\ \href@noop {} {\emph {\bibinfo
  {booktitle} {Proceedings of the 16th ACM SIGKDD International Conference on
  Knowledge Discovery and Data Mining}}},\ \bibinfo {series and number} {KDD
  '10}\ (\bibinfo  {publisher} {Association for Computing Machinery},\ \bibinfo
  {address} {New York, NY, USA},\ \bibinfo {year} {2010})\ pp.\ \bibinfo
  {pages} {1019--1028}\BibitemShut {NoStop}%
\bibitem [{\citenamefont {Kozitsin}(2023)}]{kozitzin2023b}%
  \BibitemOpen
  \bibfield  {author} {\bibinfo {author} {\bibfnamefont {I.~V.}\ \bibnamefont
  {Kozitsin}},\ }\bibfield  {title} {\enquote {\bibinfo {title} {Opinion
  formation in online public debates structured in information cascades: {A}
  system-theoretic viewpoint},}\ }\href@noop {} {\bibfield  {journal} {\bibinfo
   {journal} {Computers}\ }\textbf {\bibinfo {volume} {12}},\ \bibinfo {eid}
  {178} (\bibinfo {year} {2023})}\BibitemShut {NoStop}%
\bibitem [{\citenamefont {Ba{\~n}os}, \citenamefont {Borge-Holthoefer},\ and\
  \citenamefont {Moreno}(2013)}]{banos2013}%
  \BibitemOpen
  \bibfield  {author} {\bibinfo {author} {\bibfnamefont {R.~A.}\ \bibnamefont
  {Ba{\~n}os}}, \bibinfo {author} {\bibfnamefont {J.}~\bibnamefont
  {Borge-Holthoefer}}, \ and\ \bibinfo {author} {\bibfnamefont
  {Y.}~\bibnamefont {Moreno}},\ }\bibfield  {title} {\enquote {\bibinfo {title}
  {The role of hidden influentials in the diffusion of online information
  cascades},}\ }\href@noop {} {\bibfield  {journal} {\bibinfo  {journal}
  {European Physical Journal --- Data Science}\ }\textbf {\bibinfo {volume}
  {2}},\ \bibinfo {pages} {6} (\bibinfo {year} {2013})}\BibitemShut {NoStop}%
\bibitem [{\citenamefont {Xu}\ \emph {et~al.}(2022)\citenamefont {Xu},
  \citenamefont {Hui}, \citenamefont {Jha}, \citenamefont {Xia},\ and\
  \citenamefont {Johnson}}]{xu2022}%
  \BibitemOpen
  \bibfield  {author} {\bibinfo {author} {\bibfnamefont {C.}~\bibnamefont
  {Xu}}, \bibinfo {author} {\bibfnamefont {P.~M.}\ \bibnamefont {Hui}},
  \bibinfo {author} {\bibfnamefont {O.~K.}\ \bibnamefont {Jha}}, \bibinfo
  {author} {\bibfnamefont {C.}~\bibnamefont {Xia}}, \ and\ \bibinfo {author}
  {\bibfnamefont {N.~F.}\ \bibnamefont {Johnson}},\ }\bibfield  {title}
  {\enquote {\bibinfo {title} {Preventing the spread of online harms: {P}hysics
  of contagion across multi-platform social media and metaverses},}\
  }\href@noop {} {\bibfield  {journal} {\bibinfo  {journal} {arXiv preprint
  arXiv:2201.04249}\ } (\bibinfo {year} {2022})}\BibitemShut {NoStop}%
\bibitem [{\citenamefont {Blumberg}\ and\ \citenamefont
  {Lloyd-Smith}(2013)}]{blumberg2013}%
  \BibitemOpen
  \bibfield  {author} {\bibinfo {author} {\bibfnamefont {S.}~\bibnamefont
  {Blumberg}}\ and\ \bibinfo {author} {\bibfnamefont {J.~O.}\ \bibnamefont
  {Lloyd-Smith}},\ }\bibfield  {title} {\enquote {\bibinfo {title} {Inference
  of $r_0$ and transmission heterogeneity from the size distribution of
  stuttering chain},}\ }\href@noop {} {\bibfield  {journal} {\bibinfo
  {journal} {PLoS Computational Biology}\ }\textbf {\bibinfo {volume} {9}},\
  \bibinfo {eid} {e1002993} (\bibinfo {year} {2013})}\BibitemShut {NoStop}%
\bibitem [{Note1()}]{Note1}%
  \BibitemOpen
  \bibinfo {note} {A complementary approach is to modify opinion models by
  incorporating content~\cite {wang2024}.}\BibitemShut {Stop}%
\bibitem [{\citenamefont {Dorogovtsev}, \citenamefont {Goltsev},\ and\
  \citenamefont {Mendes}(2008)}]{dorogovtsev2008critical}%
  \BibitemOpen
  \bibfield  {author} {\bibinfo {author} {\bibfnamefont {S.~N.}\ \bibnamefont
  {Dorogovtsev}}, \bibinfo {author} {\bibfnamefont {A.~V.}\ \bibnamefont
  {Goltsev}}, \ and\ \bibinfo {author} {\bibfnamefont {J.~F.~F.}\ \bibnamefont
  {Mendes}},\ }\bibfield  {title} {\enquote {\bibinfo {title} {Critical
  phenomena in complex networks},}\ }\href@noop {} {\bibfield  {journal}
  {\bibinfo  {journal} {Reviews of Modern Physics}\ }\textbf {\bibinfo {volume}
  {80}},\ \bibinfo {pages} {1275} (\bibinfo {year} {2008})}\BibitemShut
  {NoStop}%
\bibitem [{\citenamefont {Newman}(2018)}]{newman2018networks}%
  \BibitemOpen
  \bibfield  {author} {\bibinfo {author} {\bibfnamefont {M.}~\bibnamefont
  {Newman}},\ }\href@noop {} {\emph {\bibinfo {title} {Networks}}},\ \bibinfo
  {edition} {2nd}\ ed.\ (\bibinfo  {publisher} {Oxford University Press},\
  \bibinfo {year} {2018})\BibitemShut {NoStop}%
\bibitem [{\citenamefont {Kempe}, \citenamefont {Kleinberg},\ and\
  \citenamefont {Tardos}(2003)}]{kkt2003}%
  \BibitemOpen
  \bibfield  {author} {\bibinfo {author} {\bibfnamefont {D.}~\bibnamefont
  {Kempe}}, \bibinfo {author} {\bibfnamefont {J.}~\bibnamefont {Kleinberg}}, \
  and\ \bibinfo {author} {\bibfnamefont {E.}~\bibnamefont {Tardos}},\
  }\bibfield  {title} {\enquote {\bibinfo {title} {Maximizing the spread of
  influence through a social network},}\ }in\ \href@noop {} {\emph {\bibinfo
  {booktitle} {Proceedings of the Ninth ACM SIGKDD International Conference on
  Knowledge Discovery and Data Mining}}},\ \bibinfo {series and number} {KDD
  '03}\ (\bibinfo  {publisher} {Association for Computing Machinery},\ \bibinfo
  {address} {New York, NY, USA},\ \bibinfo {year} {2003})\ pp.\ \bibinfo
  {pages} {137--146}\BibitemShut {NoStop}%
\bibitem [{\citenamefont {Vosoughi}, \citenamefont {Roy},\ and\ \citenamefont
  {Aral}(2018)}]{vosoughi2018spread}%
  \BibitemOpen
  \bibfield  {author} {\bibinfo {author} {\bibfnamefont {S.}~\bibnamefont
  {Vosoughi}}, \bibinfo {author} {\bibfnamefont {D.}~\bibnamefont {Roy}}, \
  and\ \bibinfo {author} {\bibfnamefont {S.}~\bibnamefont {Aral}},\ }\bibfield
  {title} {\enquote {\bibinfo {title} {The spread of true and false news
  online},}\ }\href@noop {} {\bibfield  {journal} {\bibinfo  {journal}
  {Science}\ }\textbf {\bibinfo {volume} {359}},\ \bibinfo {pages} {1146--1151}
  (\bibinfo {year} {2018})}\BibitemShut {NoStop}%
\bibitem [{\citenamefont {Proskurnikov}\ and\ \citenamefont
  {Tempo}(2018)}]{proskurnikov2018tutorial}%
  \BibitemOpen
  \bibfield  {author} {\bibinfo {author} {\bibfnamefont {A.~V.}\ \bibnamefont
  {Proskurnikov}}\ and\ \bibinfo {author} {\bibfnamefont {R.}~\bibnamefont
  {Tempo}},\ }\bibfield  {title} {\enquote {\bibinfo {title} {A tutorial on
  modeling and analysis of dynamic social networks. {Part II}},}\ }\href@noop
  {} {\bibfield  {journal} {\bibinfo  {journal} {Annual Reviews in Control}\
  }\textbf {\bibinfo {volume} {45}},\ \bibinfo {pages} {166--190} (\bibinfo
  {year} {2018})}\BibitemShut {NoStop}%
\bibitem [{\citenamefont {Wasserman}\ and\ \citenamefont
  {Faust}(1994)}]{faust1994}%
  \BibitemOpen
  \bibfield  {author} {\bibinfo {author} {\bibfnamefont {S.}~\bibnamefont
  {Wasserman}}\ and\ \bibinfo {author} {\bibfnamefont {K.}~\bibnamefont
  {Faust}},\ }\href@noop {} {\emph {\bibinfo {title} {Social {{Network
  Analysis}}: {{Methods}} and {{Applications}}}}}\ (\bibinfo  {publisher}
  {Cambridge University Press},\ \bibinfo {address} {Cambridge, UK},\ \bibinfo
  {year} {1994})\BibitemShut {NoStop}%
\bibitem [{\citenamefont {Bullo}(2022)}]{bullo2022}%
  \BibitemOpen
  \bibfield  {author} {\bibinfo {author} {\bibfnamefont {F.}~\bibnamefont
  {Bullo}},\ }\href@noop {} {\emph {\bibinfo {title} {Lectures on Network
  Systems}}},\ Vol.\ \bibinfo {volume} {1.6}\ (\bibinfo  {publisher} {Kindle
  Direct Publishing Santa Barbara, CA},\ \bibinfo {year} {2022})\ \bibinfo
  {note} {available at {\tt http://motion.me.ucsb.edu/book-lns/}}\BibitemShut
  {NoStop}%
\bibitem [{\citenamefont {Brooks}(2023)}]{brooks2023}%
  \BibitemOpen
  \bibfield  {author} {\bibinfo {author} {\bibfnamefont {H.~Z.}\ \bibnamefont
  {Brooks}},\ }\bibfield  {title} {\enquote {\bibinfo {title} {A tutorial on
  networks in social systems: {A} mathematical modeling perspective},}\
  }\href@noop {} {\bibfield  {journal} {\bibinfo  {journal} {arXiv preprint
  arXiv:2302.00801}\ } (\bibinfo {year} {2023})}\BibitemShut {NoStop}%
\bibitem [{\citenamefont {Lerman}\ and\ \citenamefont
  {Ghosh}(2010)}]{lerman2010information}%
  \BibitemOpen
  \bibfield  {author} {\bibinfo {author} {\bibfnamefont {K.}~\bibnamefont
  {Lerman}}\ and\ \bibinfo {author} {\bibfnamefont {R.}~\bibnamefont {Ghosh}},\
  }\bibfield  {title} {\enquote {\bibinfo {title} {Information contagion: {A}n
  empirical study of the spread of news on {D}igg and {T}witter social
  networks},}\ }in\ \href@noop {} {\emph {\bibinfo {booktitle} {Proceedings of
  the Fourth International AAAI Conference on Weblogs and Social Media}}},\
  \bibinfo {series and number} {ICWSM-10}\ (\bibinfo  {publisher} {Association
  for the Advancement of Artificial Intelligence},\ \bibinfo {address}
  {Washington, DC, USA},\ \bibinfo {year} {2010})\BibitemShut {NoStop}%
\bibitem [{\citenamefont {Gleeson}\ \emph {et~al.}(2014)\citenamefont
  {Gleeson}, \citenamefont {Ward}, \citenamefont {O'Sullivan},\ and\
  \citenamefont {Lee}}]{gleeson2014}%
  \BibitemOpen
  \bibfield  {author} {\bibinfo {author} {\bibfnamefont {J.~P.}\ \bibnamefont
  {Gleeson}}, \bibinfo {author} {\bibfnamefont {J.~A.}\ \bibnamefont {Ward}},
  \bibinfo {author} {\bibfnamefont {K.~P.}\ \bibnamefont {O'Sullivan}}, \ and\
  \bibinfo {author} {\bibfnamefont {W.~T.}\ \bibnamefont {Lee}},\ }\bibfield
  {title} {\enquote {\bibinfo {title} {Competition-induced criticality in a
  model of meme popularity},}\ }\href@noop {} {\bibfield  {journal} {\bibinfo
  {journal} {Physical Review Letters}\ }\textbf {\bibinfo {volume} {112}},\
  \bibinfo {eid} {048701} (\bibinfo {year} {2014})}\BibitemShut {NoStop}%
\bibitem [{\citenamefont {Adamic}\ \emph {et~al.}(2016)\citenamefont {Adamic},
  \citenamefont {Lento}, \citenamefont {Adar},\ and\ \citenamefont
  {Ng}}]{adamic2016}%
  \BibitemOpen
  \bibfield  {author} {\bibinfo {author} {\bibfnamefont {L.~A.}\ \bibnamefont
  {Adamic}}, \bibinfo {author} {\bibfnamefont {T.~M.}\ \bibnamefont {Lento}},
  \bibinfo {author} {\bibfnamefont {E.}~\bibnamefont {Adar}}, \ and\ \bibinfo
  {author} {\bibfnamefont {P.~C.}\ \bibnamefont {Ng}},\ }\bibfield  {title}
  {\enquote {\bibinfo {title} {Information evolution in social networks},}\
  }in\ \href@noop {} {\emph {\bibinfo {booktitle} {Proceedings of the Ninth ACM
  International Conference on Web Search and Data Mining}}},\ \bibinfo {series
  and number} {WSDM '16}\ (\bibinfo  {publisher} {Association for Computing
  Machinery},\ \bibinfo {address} {New York, NY, USA},\ \bibinfo {year}
  {2016})\ pp.\ \bibinfo {pages} {473--482}\BibitemShut {NoStop}%
\bibitem [{\citenamefont {Oh}\ and\ \citenamefont
  {Porter}(2018)}]{oh2018complex}%
  \BibitemOpen
  \bibfield  {author} {\bibinfo {author} {\bibfnamefont {S.-W.}\ \bibnamefont
  {Oh}}\ and\ \bibinfo {author} {\bibfnamefont {M.~A.}\ \bibnamefont
  {Porter}},\ }\bibfield  {title} {\enquote {\bibinfo {title} {Complex
  contagions with timers},}\ }\href@noop {} {\bibfield  {journal} {\bibinfo
  {journal} {Chaos: An Interdisciplinary Journal of Nonlinear Science}\
  }\textbf {\bibinfo {volume} {28}},\ \bibinfo {eid} {033101} (\bibinfo {year}
  {2018})}\BibitemShut {NoStop}%
\bibitem [{\citenamefont {Min}\ and\ \citenamefont
  {San~Miguel}(2018)}]{min2018competing}%
  \BibitemOpen
  \bibfield  {author} {\bibinfo {author} {\bibfnamefont {B.}~\bibnamefont
  {Min}}\ and\ \bibinfo {author} {\bibfnamefont {M.}~\bibnamefont
  {San~Miguel}},\ }\bibfield  {title} {\enquote {\bibinfo {title} {Competing
  contagion processes: {C}omplex contagion triggered by simple contagion},}\
  }\href@noop {} {\bibfield  {journal} {\bibinfo  {journal} {Scientific
  Reports}\ }\textbf {\bibinfo {volume} {8}},\ \bibinfo {pages} {10422}
  (\bibinfo {year} {2018})}\BibitemShut {NoStop}%
\bibitem [{\citenamefont {Newman}, \citenamefont {Strogatz},\ and\
  \citenamefont {Watts}(2001)}]{newman2001random}%
  \BibitemOpen
  \bibfield  {author} {\bibinfo {author} {\bibfnamefont {M.~E.~J.}\
  \bibnamefont {Newman}}, \bibinfo {author} {\bibfnamefont {S.~H.}\
  \bibnamefont {Strogatz}}, \ and\ \bibinfo {author} {\bibfnamefont {D.~J.}\
  \bibnamefont {Watts}},\ }\bibfield  {title} {\enquote {\bibinfo {title}
  {Random graphs with arbitrary degree distributions and their applications},}\
  }\href@noop {} {\bibfield  {journal} {\bibinfo  {journal} {Physical Review
  E}\ }\textbf {\bibinfo {volume} {64}},\ \bibinfo {pages} {026118} (\bibinfo
  {year} {2001})}\BibitemShut {NoStop}%
\bibitem [{\citenamefont {Newman}(2002)}]{newman2002spread}%
  \BibitemOpen
  \bibfield  {author} {\bibinfo {author} {\bibfnamefont {M.~E.~J.}\
  \bibnamefont {Newman}},\ }\bibfield  {title} {\enquote {\bibinfo {title}
  {Spread of epidemic disease on networks},}\ }\href@noop {} {\bibfield
  {journal} {\bibinfo  {journal} {Physical Review E}\ }\textbf {\bibinfo
  {volume} {66}},\ \bibinfo {pages} {016128} (\bibinfo {year}
  {2002})}\BibitemShut {NoStop}%
\bibitem [{\citenamefont {No{\"e}l}\ \emph {et~al.}(2009)\citenamefont
  {No{\"e}l}, \citenamefont {Davoudi}, \citenamefont {Brunham}, \citenamefont
  {Dub{\'e}},\ and\ \citenamefont {Pourbohloul}}]{noel2009time}%
  \BibitemOpen
  \bibfield  {author} {\bibinfo {author} {\bibfnamefont {P.-A.}\ \bibnamefont
  {No{\"e}l}}, \bibinfo {author} {\bibfnamefont {B.}~\bibnamefont {Davoudi}},
  \bibinfo {author} {\bibfnamefont {R.~C.}\ \bibnamefont {Brunham}}, \bibinfo
  {author} {\bibfnamefont {L.~J.}\ \bibnamefont {Dub{\'e}}}, \ and\ \bibinfo
  {author} {\bibfnamefont {B.}~\bibnamefont {Pourbohloul}},\ }\bibfield
  {title} {\enquote {\bibinfo {title} {Time evolution of epidemic disease on
  finite and infinite networks},}\ }\href@noop {} {\bibfield  {journal}
  {\bibinfo  {journal} {Physical Review E}\ }\textbf {\bibinfo {volume} {79}},\
  \bibinfo {eid} {026101} (\bibinfo {year} {2009})}\BibitemShut {NoStop}%
\bibitem [{\citenamefont {Fosdick}\ \emph {et~al.}(2018)\citenamefont
  {Fosdick}, \citenamefont {Larremore}, \citenamefont {Nishimura},\ and\
  \citenamefont {Ugander}}]{fosdick2018}%
  \BibitemOpen
  \bibfield  {author} {\bibinfo {author} {\bibfnamefont {B.~K.}\ \bibnamefont
  {Fosdick}}, \bibinfo {author} {\bibfnamefont {D.~B.}\ \bibnamefont
  {Larremore}}, \bibinfo {author} {\bibfnamefont {J.}~\bibnamefont
  {Nishimura}}, \ and\ \bibinfo {author} {\bibfnamefont {J.}~\bibnamefont
  {Ugander}},\ }\bibfield  {title} {\enquote {\bibinfo {title} {Configuring
  random graph models with fixed degree sequences},}\ }\href@noop {} {\bibfield
   {journal} {\bibinfo  {journal} {SIAM Review}\ }\textbf {\bibinfo {volume}
  {60}},\ \bibinfo {pages} {315--355} (\bibinfo {year} {2018})}\BibitemShut
  {NoStop}%
\bibitem [{\citenamefont {Peng}\ \emph {et~al.}(2018)\citenamefont {Peng},
  \citenamefont {Zhou}, \citenamefont {Cao}, \citenamefont {Yu}, \citenamefont
  {Niu},\ and\ \citenamefont {Jia}}]{peng2018}%
  \BibitemOpen
  \bibfield  {author} {\bibinfo {author} {\bibfnamefont {S.}~\bibnamefont
  {Peng}}, \bibinfo {author} {\bibfnamefont {Y.}~\bibnamefont {Zhou}}, \bibinfo
  {author} {\bibfnamefont {L.}~\bibnamefont {Cao}}, \bibinfo {author}
  {\bibfnamefont {S.}~\bibnamefont {Yu}}, \bibinfo {author} {\bibfnamefont
  {J.}~\bibnamefont {Niu}}, \ and\ \bibinfo {author} {\bibfnamefont
  {W.}~\bibnamefont {Jia}},\ }\bibfield  {title} {\enquote {\bibinfo {title}
  {Influence analysis in social networks: {A} survey},}\ }\href@noop {}
  {\bibfield  {journal} {\bibinfo  {journal} {Journal of Network and Computer
  Applications}\ }\textbf {\bibinfo {volume} {106}},\ \bibinfo {pages} {17--32}
  (\bibinfo {year} {2018})}\BibitemShut {NoStop}%
\bibitem [{\citenamefont {She}, \citenamefont {Par\'{e}},\ and\ \citenamefont
  {Hale}(2023)}]{she2023}%
  \BibitemOpen
  \bibfield  {author} {\bibinfo {author} {\bibfnamefont {B.}~\bibnamefont
  {She}}, \bibinfo {author} {\bibfnamefont {P.~E.}\ \bibnamefont {Par\'{e}}}, \
  and\ \bibinfo {author} {\bibfnamefont {M.}~\bibnamefont {Hale}},\ }\bibfield
  {title} {\enquote {\bibinfo {title} {Distributed reproduction numbers of
  networked epidemics},}\ }in\ \href@noop {} {\emph {\bibinfo {booktitle} {IEEE
  Control Systems Letters}}},\ \bibinfo {series and number} {American Control
  Conference (ACC)}\ (\bibinfo  {publisher} {Institute of Electrical and
  Electronics Engineers},\ \bibinfo {address} {San Diego, CA, USA},\ \bibinfo
  {year} {2023})\BibitemShut {NoStop}%
\bibitem [{\citenamefont {Melnik}\ \emph {et~al.}(2014)\citenamefont {Melnik},
  \citenamefont {Porter}, \citenamefont {Mucha},\ and\ \citenamefont
  {Gleeson}}]{melnik2014}%
  \BibitemOpen
  \bibfield  {author} {\bibinfo {author} {\bibfnamefont {S.}~\bibnamefont
  {Melnik}}, \bibinfo {author} {\bibfnamefont {M.~A.}\ \bibnamefont {Porter}},
  \bibinfo {author} {\bibfnamefont {P.~J.}\ \bibnamefont {Mucha}}, \ and\
  \bibinfo {author} {\bibfnamefont {J.~P.}\ \bibnamefont {Gleeson}},\
  }\bibfield  {title} {\enquote {\bibinfo {title} {Dynamics on modular networks
  with heterogeneous correlations},}\ }\href@noop {} {\bibfield  {journal}
  {\bibinfo  {journal} {Chaos: An Interdisciplinary Journal of Nonlinear
  Science}\ }\textbf {\bibinfo {volume} {24}},\ \bibinfo {eid} {023106}
  (\bibinfo {year} {2014})}\BibitemShut {NoStop}%
\bibitem [{\citenamefont {Kivel{\"a}}\ \emph {et~al.}(2014)\citenamefont
  {Kivel{\"a}}, \citenamefont {Arenas}, \citenamefont {Barthelemy},
  \citenamefont {Gleeson}, \citenamefont {Moreno},\ and\ \citenamefont
  {Porter}}]{kivela2014}%
  \BibitemOpen
  \bibfield  {author} {\bibinfo {author} {\bibfnamefont {M.}~\bibnamefont
  {Kivel{\"a}}}, \bibinfo {author} {\bibfnamefont {A.}~\bibnamefont {Arenas}},
  \bibinfo {author} {\bibfnamefont {M.}~\bibnamefont {Barthelemy}}, \bibinfo
  {author} {\bibfnamefont {J.~P.}\ \bibnamefont {Gleeson}}, \bibinfo {author}
  {\bibfnamefont {Y.}~\bibnamefont {Moreno}}, \ and\ \bibinfo {author}
  {\bibfnamefont {M.~A.}\ \bibnamefont {Porter}},\ }\bibfield  {title}
  {\enquote {\bibinfo {title} {Multilayer networks},}\ }\href@noop {}
  {\bibfield  {journal} {\bibinfo  {journal} {Journal of Complex Networks}\
  }\textbf {\bibinfo {volume} {2}},\ \bibinfo {pages} {203--271} (\bibinfo
  {year} {2014})}\BibitemShut {NoStop}%
\bibitem [{\citenamefont {Battiston}\ \emph {et~al.}(2020)\citenamefont
  {Battiston}, \citenamefont {Cencetti}, \citenamefont {Iacopini},
  \citenamefont {Latora}, \citenamefont {Lucas}, \citenamefont {Patania},
  \citenamefont {Young},\ and\ \citenamefont {Petri}}]{battiston2020}%
  \BibitemOpen
  \bibfield  {author} {\bibinfo {author} {\bibfnamefont {F.}~\bibnamefont
  {Battiston}}, \bibinfo {author} {\bibfnamefont {G.}~\bibnamefont {Cencetti}},
  \bibinfo {author} {\bibfnamefont {I.}~\bibnamefont {Iacopini}}, \bibinfo
  {author} {\bibfnamefont {V.}~\bibnamefont {Latora}}, \bibinfo {author}
  {\bibfnamefont {M.}~\bibnamefont {Lucas}}, \bibinfo {author} {\bibfnamefont
  {A.}~\bibnamefont {Patania}}, \bibinfo {author} {\bibfnamefont {J.-G.}\
  \bibnamefont {Young}}, \ and\ \bibinfo {author} {\bibfnamefont
  {G.}~\bibnamefont {Petri}},\ }\bibfield  {title} {\enquote {\bibinfo {title}
  {Networks beyond pairwise interactions: {S}tructure and dynamics},}\
  }\href@noop {} {\bibfield  {journal} {\bibinfo  {journal} {Physics Reports}\
  }\textbf {\bibinfo {volume} {874}},\ \bibinfo {pages} {1--92} (\bibinfo
  {year} {2020})}\BibitemShut {NoStop}%
\bibitem [{\citenamefont {Berner}\ \emph {et~al.}(2023)\citenamefont {Berner},
  \citenamefont {Gross}, \citenamefont {Kuehn}, \citenamefont {Kurths},\ and\
  \citenamefont {Yanchuk}}]{berner2023}%
  \BibitemOpen
  \bibfield  {author} {\bibinfo {author} {\bibfnamefont {R.}~\bibnamefont
  {Berner}}, \bibinfo {author} {\bibfnamefont {T.}~\bibnamefont {Gross}},
  \bibinfo {author} {\bibfnamefont {C.}~\bibnamefont {Kuehn}}, \bibinfo
  {author} {\bibfnamefont {J.}~\bibnamefont {Kurths}}, \ and\ \bibinfo {author}
  {\bibfnamefont {S.}~\bibnamefont {Yanchuk}},\ }\bibfield  {title} {\enquote
  {\bibinfo {title} {Adaptive dynamical networks},}\ }\href@noop {} {\bibfield
  {journal} {\bibinfo  {journal} {Physics Reports}\ }\textbf {\bibinfo {volume}
  {1031}},\ \bibinfo {pages} {1--59} (\bibinfo {year} {2023})}\BibitemShut
  {NoStop}%
\bibitem [{\citenamefont {Deffuant}\ \emph {et~al.}(2000)\citenamefont
  {Deffuant}, \citenamefont {Neau}, \citenamefont {Amblard},\ and\
  \citenamefont {Weisbuch}}]{deffuant2000mixing}%
  \BibitemOpen
  \bibfield  {author} {\bibinfo {author} {\bibfnamefont {G.}~\bibnamefont
  {Deffuant}}, \bibinfo {author} {\bibfnamefont {D.}~\bibnamefont {Neau}},
  \bibinfo {author} {\bibfnamefont {F.}~\bibnamefont {Amblard}}, \ and\
  \bibinfo {author} {\bibfnamefont {G.}~\bibnamefont {Weisbuch}},\ }\bibfield
  {title} {\enquote {\bibinfo {title} {Mixing beliefs among interacting
  agents},}\ }\href@noop {} {\bibfield  {journal} {\bibinfo  {journal}
  {Advances in Complex Systems}\ }\textbf {\bibinfo {volume} {3}},\ \bibinfo
  {pages} {87--98} (\bibinfo {year} {2000})}\BibitemShut {NoStop}%
\bibitem [{\citenamefont {Porter}(2002)}]{porter2002}%
  \BibitemOpen
  \bibfield  {author} {\bibinfo {author} {\bibfnamefont {M.~A.}\ \bibnamefont
  {Porter}},\ }\href@noop {} {\enquote {\bibinfo {title} {{Quantum Chaos in
  Vibrating Billiard Systems}},}\ } (\bibinfo {year} {2002}),\ \bibinfo {note}
  {{Ph.D. thesis, Cornell University}}\BibitemShut {NoStop}%
\bibitem [{\citenamefont {Porter}\ \emph {et~al.}(2005)\citenamefont {Porter},
  \citenamefont {Carretero-Gonz\'{a}lez}, \citenamefont {Kevrekidis},\ and\
  \citenamefont {Maolmed}}]{porter2005}%
  \BibitemOpen
  \bibfield  {author} {\bibinfo {author} {\bibfnamefont {M.~A.}\ \bibnamefont
  {Porter}}, \bibinfo {author} {\bibfnamefont {R.}~\bibnamefont
  {Carretero-Gonz\'{a}lez}}, \bibinfo {author} {\bibfnamefont {P.~G.}\
  \bibnamefont {Kevrekidis}}, \ and\ \bibinfo {author} {\bibfnamefont {B.~A.}\
  \bibnamefont {Maolmed}},\ }\bibfield  {title} {\enquote {\bibinfo {title}
  {Nonlinear lattice dynamics of {B}ose--{E}instein condensates},}\ }\href@noop
  {} {\bibfield  {journal} {\bibinfo  {journal} {Chaos: An Interdisciplinary
  Journal of Nonlinear Science}\ }\textbf {\bibinfo {volume} {15}},\ \bibinfo
  {eid} {015115} (\bibinfo {year} {2005})}\BibitemShut {NoStop}%
\bibitem [{\citenamefont {Porter}, \citenamefont {Zabusky},\ and\ \citenamefont
  {Campbell}(2009)}]{porter2009}%
  \BibitemOpen
  \bibfield  {author} {\bibinfo {author} {\bibfnamefont {M.~A.}\ \bibnamefont
  {Porter}}, \bibinfo {author} {\bibfnamefont {N.~J.}\ \bibnamefont {Zabusky}},
  \ and\ \bibinfo {author} {\bibfnamefont {D.~K.}\ \bibnamefont {Campbell}},\
  }\bibfield  {title} {\enquote {\bibinfo {title} {{F}ermi, {P}asta, {U}lam and
  the birth of experimental mathematics},}\ }\href@noop {} {\bibfield
  {journal} {\bibinfo  {journal} {American Scientist}\ }\textbf {\bibinfo
  {volume} {97}},\ \bibinfo {pages} {214--221} (\bibinfo {year}
  {2009})}\BibitemShut {NoStop}%
\bibitem [{\citenamefont {Nelson}, \citenamefont {Porter},\ and\ \citenamefont
  {Choubey}(2018)}]{nelson2018}%
  \BibitemOpen
  \bibfield  {author} {\bibinfo {author} {\bibfnamefont {H.}~\bibnamefont
  {Nelson}}, \bibinfo {author} {\bibfnamefont {M.~A.}\ \bibnamefont {Porter}},
  \ and\ \bibinfo {author} {\bibfnamefont {B.}~\bibnamefont {Choubey}},\
  }\bibfield  {title} {\enquote {\bibinfo {title} {Variability in
  {F}ermi--{P}asta--{U}lam--{T}singou arrays can prevent recurrences},}\
  }\href@noop {} {\bibfield  {journal} {\bibinfo  {journal} {Physical Review
  E}\ }\textbf {\bibinfo {volume} {98}},\ \bibinfo {eid} {062210} (\bibinfo
  {year} {2018})}\BibitemShut {NoStop}%
\bibitem [{\citenamefont {Wang}, \citenamefont {Chen},\ and\ \citenamefont
  {Zhao}(2024)}]{wang2024}%
  \BibitemOpen
  \bibfield  {author} {\bibinfo {author} {\bibfnamefont {H.}~\bibnamefont
  {Wang}}, \bibinfo {author} {\bibfnamefont {Z.}~\bibnamefont {Chen}}, \ and\
  \bibinfo {author} {\bibfnamefont {H.~V.}\ \bibnamefont {Zhao}},\ }\bibfield
  {title} {\enquote {\bibinfo {title} {Message-enhanced {DeGroot} model},}\
  }\href@noop {} {\bibfield  {journal} {\bibinfo  {journal} {arXiv preprint
  arXiv:2402.18867}\ } (\bibinfo {year} {2024})}\BibitemShut {NoStop}%
\end{thebibliography}


%


\end{document}